\documentclass[aps,prb,twocolumn,superscriptaddress]{revtex4-1}
\usepackage{amsmath,amssymb}
\usepackage{graphicx}
\usepackage{subfigure}
\usepackage{verbatim}
\usepackage{amsfonts}
\usepackage{dcolumn}
\usepackage{bm}
\usepackage{color}
\usepackage[colorlinks,citecolor=blue]{hyperref}

\usepackage{mathrsfs}

\newcommand{\beq}{\begin{eqnarray} }
	\newcommand{\eeq}{\end{eqnarray} }
\newcommand{\Beq}{\begin{eqnarray*} }
	\newcommand{\Eeq}{\end{eqnarray*} }
\newcommand{\Bmat}{\left(\begin{matrix}}
	\newcommand{\Emat}{\end{matrix}\right)}
\newcommand{\up}{\uparrow}
\newcommand{\dn}{\downarrow}

\begin{document}
	
\title{Realization of the topological Hopf term in two-dimensional  lattice models}
	
\author{Yan-Guang Yue}
\affiliation{Department of Physics, Renmin University of China, Beijing 100872, China}
	
\author{Zheng-Xin Liu}
\email{liuzxphys@ruc.edu.cn}
\affiliation{Department of Physics, Renmin University of China, Beijing 100872, China}
	
\date{\today}
	
\begin{abstract}

It is known that a two-dimensional spin system can acquire a topological Hopf term by coupling to massless Dirac fermions whose energy spectrum has a single cone. But it is challenging to realize the Hopf term in condensed matter physics due to the fermion-doubling in the low-energy spectrum. In this work we propose a scenario to realize the Hopf term in lattice models. The central aim is tuning the coupling between the spins and the Dirac fermions such that the topological terms contributed by the two cones do not cancel each other. To this end, we consider $p_x$ and $p_y$ orbitals for the Dirac fermions on the honeycomb lattice such that there are totally four bands.  
By utilizing the orbital degrees of freedom, a $\theta=2\pi$ Hopf term is successfully generated for the spin system after integrating out the Dirac fermions. If the fermions have a small gap $m_0$ or if the spin-orbit coupling is considered, then $\theta$ is no longer quantized, but it may flow to multiple of $2\pi$ under renormalization. The ground state and the physical response of a spin system having the Hopf term are discussed. 
\end{abstract}
	
\maketitle
	
\section{ Introduction} 
Most topological phases can be described by topological terms in their path-integral. In continuous quantum field theory, the topological terms do not depend on the metric of space-time. For instance, integer and fractional quantum Hall liquids\cite{ZHK_CS_FQH, LopezFradkin_CS_FQH} and chiral spin liquids\cite{KL_CSL, WenWilczekZee_CSL, Wen_CSL} are described by Chern-Simons terms in the effective gauge theory. On the other hand, the Haldane phase for a spin chain with integer spin is described by a topological $\theta$-term\cite{HaldanePLA1983, HaldanePRL1983} which is quantized to an integer times $2\pi$ if the space-time manifold is closed. Topological $\theta$-terms can also be used to describe topological insulators \cite{QiHughesZhang_TI}. Systems having topological terms in their path-integral can also be gapless. A typical example is the spin-1/2 antiferromagnetic Heisenberg chain which has $\pi$-quantized $\theta$-term or Wess-Zumino-Witten term in its Lagrangian\cite{WZW}. Consequently, the system is gapless and respects the Lieb-Schultz-Mattis theorem\cite{LSM}.  Among the gapped phases, 
some have nontrivial topological orders which are characterized by fractional excitations (called anyons) or chiral edge states, but some don't have anyonic excitations and contain no topological order. However, the nontrivial topological terms indicate that if certain symmetry is present, gapped systems with trivial topological orders can still have protected edge states. 
The ground states of these systems are known as the symmetry protected topological (SPT) states \cite{GuWen2009,  ChenGuLiuWen2011, ChenGuLiuWenSCI} which are adiabatically connected to a trivial state if the protecting symmetry is explicitly broken. 

The simplest Bosonic SPT phase is the spin-$1$ Haldane phase. Although one-dimensional SPT phases are more precisely described by projective representations and classified by the second group cohomology of the symmetry group\cite{ChenGuWen2011_1D, ChenGuWen2011_1Dfull}, the Haldane chain was originally understood from the topological field theory,  namely (1+1)-dimensional $SO(3)$ nonlinear sigma model (NLSM) with topological $\theta$-term, i.e. ,
	$
	S_{\rm nlsm} =\int dxdt \big[(\partial_t \pmb n)^2 - (\partial_x \pmb n)^2\big] + S_{\theta} 
	$,
	where $\partial_t, \partial_x$ denote ${\partial\over\partial t}$ and ${\partial\over\partial x}$ ,respectively, $\pmb n(x,t)$ is a vector field in space-time, and
	$$
	 S_{\theta} = {\theta\over4\pi} \int dxdt \pmb n\cdot {\partial_t\pmb n}\times{\partial_x \pmb n}.
	$$
	The above effective action  can be  derived from the microscopic lattice model — the spin-$1$ Heisenberg model ($\theta=2\pi $). If space-time is closed, then the $\theta$-term quantizes to $2\pi S$ times the skyrmion number $N$ of the spin configuration in the space-time manifold.  The integer $N\in \pi_2(S^2)=\mathbb Z$ is essentially the mapping degree from space-time manifold $S^2$ to the symmetric space of the $SO(3)$ group (which is also $S^2$). When summing over all the configurations with different skyrmion numbers,  the system becomes short-range correlated and gapped. If the system has a boundary, then the $\theta$-term $e^{iS_\theta}$ can be identified with the Berry phase of the spin-${1\over2}$ edge states.\cite{Ng1994} Therefore,  the NLSM with $\theta$-term (called topological NLSM) successfully describes the Haldane phase.

	The $SO(3)$ topological NLSM was generalized to higher dimensions to describe and classify general bosonic SPT phases\cite{Xu2013}. In $d$-spacial dimensions, associated with the dynamical term of $SO(d+2)$ NLSM, one can construct a $\theta$-term, 
	$$
	 S_{\theta}={2\pi \mathcal K\over V}\int d^{d} xdt \epsilon^{\mu_1...\mu_d}\epsilon^{i_1...i_d} \big( n^{i_1} \partial_{{\mu_1} } n^{i_2} ... \partial_{{\mu_{d}}} n^{i_{d+1}}\big)
	$$ 
	where $V$ is the volume of the sphere $S^{d+1}$ and  $ {S_{\theta}\over(2\pi K)}\in \pi_{d+1}(S^{d+1})=\mathbb Z$ is essentially the mapping degree from the space-time manifold $S^{d+1}$ to the symmetric space $S^{d+1}$ of the group $SO(d+2)$. This $SO(d+2)$ NLSM can be used to describe and classify bosonic SPT phases whose symmetry group is a certain subgroup of $SO(d+2)\times Z_2^T$, where $Z_2^T$ stands  for the time-reversal symmetry \cite{Xu2013}.

In (2+1)-dimensions, a special kind of $\theta$-term is the Hopf term [see Eqs.(\ref{hopf1}) and (\ref{Hopfind})] in $SO(3)$ NLSM. The Hopf term originates from the Hopf map from the space-time manifold $S^3$ to the symmetric space $S^2$, which has $\pi_3(S^2)=\mathbb Z$ distinct topological classes. The Hopf term can change the statistics of the 
skyrmions\cite{WilczekZee, WuZee}. Hence if a spin system contains a Hopf term in its Lagrangian and if its ground state is gapped without spontaneous symmetry breaking, then it may contain either intrinsic topological order\cite{WenNiu90, Wen1990} or symmetry protected topological order\cite{LiuWen}. A possible way to obtain the Hopf term is coupling the spins to Dirac fermions\cite{Abanov, AbanovWiegmann, Huan}.
Integrating out the fermions gives rise to a $\theta=\pi$ Hopf term for the spins which cannot be obtained in a perturbative way. A consequence of the $\pi$-quantized Hopf term is that a skyrmion traps a fermion in its core and carries 1/2 angular momentum\cite{WilczekZee, Huan}.  
	
However, lattice model realizing the Hopf term in their Lagrangian is still lacking. Owing to the fermion doubling theorem, a single Dirac cone cannot be obtained in lattice models without fine-tuning.

One should couple the spin system to at least one pair of Dirac cones. However, 
a straightforward coupling essentially results in a cancellation in the topological term. Therefore, one needs to introduce more degrees of freedom for
cancelation. In the rest part of the paper, we
illustrate that a Hopf term can be obtained by introducing $p_x, p_y$ orbitals to the fermions. We  further show that the resultant ground state belongs to a $SU(2)$ symmetry protected SPT phase. This provides a possible scheme to obtain topological terms for other symmetry groups and then to prepare for the corresponding SPT phases. 
	
The rest part of the paper is organized as the following. In Sec.\ref{sec:gapless} we discuss in detail the scenario of obtaining a $\theta=2\pi$ Hopf term by coupling spins to lattice Dirac fermions. The physical consequences of the Hopf term are discussed in Sec.\ref{sec:spt}. Since the Dirac cones may have a small gap in real materials, in Sec.\ref{sec:gapped} we discuss the effect of the mass or spin-orbit coupling for the Dirac fermions.
Section.\ref{sec:summary} is devoted to the conclusions and discussions. 
	
\section{Hopf model From massless Dirac fermions}	\label{sec:gapless}
\subsection{Continuum model: A single Dirac cone} 
Firstly, we briefly review the scenario proposed by Abanov and Wiegmann \cite{Abanov, AbanovWiegmann} on the appearance of the Hopf term in the Lagrangian of spin systems by coupling to massless Dirac fermions. Suppose there are massless spin-1/2 fermions $\psi(\pmb r)$ forming a dispersion with a single Dirac cone $E_{\pmb k} = |\pmb k|$ in the momentum space. The action of the massless Dirac fermions is 
\Beq
S_0 = \int  d^{2}xdt \bar{\psi} i\gamma^{\mu}\partial_{\mu} \psi, 
\Eeq
where $\{\gamma^\mu, \gamma^\nu\}=2 \mathfrak  g ^{\mu\nu}$ with $\mathfrak g={\rm diag}(1,-1,-1)$ the space-time metric and $\bar\psi =\psi^\dag \gamma_0$. From now on we will replace $\int d^2xdt$ with $\int d^3x$. Then we couple the fermions to a two-component spin field $z=(z_1, z_2)^T$ via 
$$
S_{\rm int}=m\bar\psi \pmb \sigma\psi \cdot \pmb n(x), 
$$ 
with $\pmb n=z^\dag \pmb \sigma z$, $z^\dag z=1$ and $m$ the magnitude of the spin which can be considered as a constant. If the spin field $\pmb n(x)$ is uniform in space-time, then the Dirac fermions obtain a finite mass and open a gap with a dispersion $E_{\pmb k}=\sqrt{m^2 + k^2}$. 
	
	If the vector field $\pmb n(x)$ is a smooth function of space and time, then integrating out the fermions will result in a Hopf term for the spins (see Appendix \ref{app:singlecone})
	\beq\label{hopf1}
	 S_{\rm hopf} =  \pi\ {\rm sgn}(m) H(\pmb n),
	\eeq
	with 
	\beq\label{Hopfind}
	H(\pmb n) = -{1\over 4\pi^2}\int d^3x \epsilon^{\mu\nu\lambda} a_\mu\partial_\nu a_\lambda 
	\eeq
	the Hopf invariant for the mapping from the space-time manifold $S^3$ to the $S^2$ formed by $\pmb n$. Here $a_\mu = -i z^\dag \partial_{\mu} z$ where $z$ is the eigenstate of $\pmb n\cdot\pmb \sigma$ known as the spin coherent state. The Hopf invariant  $H(\pmb n)$ is nonlocal in forms of $\pmb n$ (later we will give an expression in forms of local variables). The derivation of the above Hopf term is subtle because Eq.(\ref{hopf1}) is invariant under small variations of the field $\pmb n(x)$ thus it cannot be obtained in a perturbative way. Besides the Hopf term, a dynamical $SO(3)$ NLSM term is also obtained by integrating out the fermions. Therefore, one obtains the (2+1)-dimensional $SO(3)$ topological NLSM
	\beq\label{O3nlsm}
	 S_{\rm nlsm} = \int d^{3}x \Big[{1\over\lambda} \Big((\partial_t\pmb n)^2 - (\nabla \pmb n)^2\Big)  + \theta H(\pmb n) \Big],
	\eeq
	with $\theta = \pi\ {\rm sgn}(m)$  and $\lambda ={8\pi\over |m|}$. In later discussions, we will  call the above model as the `Hopf (NLSM) model'.

\subsection{Lattice model: A pair of Dirac cones} 
	
Now we try to realize the Hopf model (\ref{O3nlsm}) via microscopic lattice models. 
We should first introduce the massless Dirac fermions.  As a well known candidate, the graphene \cite{Graphene} with the honeycomb lattice structure can host massless Dirac fermions in the low energy limit. Owing to the fermion doubling, a couple of Dirac cones, called valleys, appear at the fermi level under half filling. However, as shown below,  straightforwardly adding the contribution of the two cones results in a cancelation in the topological term. One needs to seek for more degrees of freedom and to carefully design the coupling between the fermions and the spins. 
We then provide an accessible scenario.

	{\bf No Hopf term from graphene.} The tight binding Hamiltonian for electrons (with a single $p_z$-orbital) in the graphene  reads $H_0=\sum_{\langle i,j\rangle,\sigma=\up,\dn} (t c_{i\sigma}^\dag c_{j\sigma} + {\rm h.c.})$, where only the nearest neighbor hoppings are considered. The two sublattices are labeled by $A$ and $B$, and the lattice translation group is generated by $\pmb a_1=\hat x, \pmb a_2 ={1\over2}\hat x+{\sqrt3\over2}\hat y$. Accordingly, the bases in the reciprocal lattice are given by $\pmb b_1,\pmb b_2$ with $\pmb a_i\cdot\pmb b_j=2\pi\delta_{ij}$. When diagonalizing the Hamiltonian in momentum space, one finds two Dirac cones in the energy bands locating at $\pmb K={1\over3}\pmb b_1 + {2\over3}\pmb b_2=({2\pi\over3},{2\sqrt3\pi\over3})$ and $\pmb K'={2\over3}\pmb b_1 + {1\over3}\pmb b_2=({4\pi\over3},0)$. The effective Hamiltonian at $K$ and $K'$ are given by $H_{K+k} = v(k_x\gamma^x + k_y\gamma^y)$ and $H_{K'+k} = v(-k_x\gamma^x + k_y\gamma^y)$, respectively, where $v$ is the fermi velocity and $\pmb k=(k_x,k_y)$ is a small relative momentum. Introducing the low-energy bases $\psi^\dag (x)=[\psi^\dag_{KA}(x),\psi^\dag_{KB}(x), \psi^\dag_{K'A}(x),\psi^\dag_{K'B}(x) ]$, then the effective theory for the graphene is described by 
$$
S_{\rm grph} = \int d^3x [i\psi^\dag \partial_t \psi + i\psi^\dag (\mathcal V_z\gamma^x\partial_x +  \gamma^y\partial_y)\psi],
$$
where $\gamma^{x,y,z}$ are three Pauli matrices acting on the sublattice indices $A,B$ and $\mathcal V_z={\rm diag}(1, -1)$ acts on  the valley indices $K$ and $K'$. The fermi velocity has been rescaled as $v=1$. Denoting $\gamma^0=-\gamma^z,  \gamma^1=  \gamma^0\gamma^x, \gamma^2= \gamma^0\gamma^y$ with $(\gamma^0)^2=1, (\gamma^{1})^2=(\gamma^2)^2=-1$, the above action can be written in a relativistic form 
\Beq
S_{\rm grph} =\int d^3x \bar\psi (i \gamma^0\partial_0 + i\mathcal V_z\gamma^1\partial_1 + i\gamma^2 \partial_2 ) \psi, 
\Eeq
where $\partial_{0,1,2}\equiv \partial_{t,x,y}$ and $\bar\psi=\psi^\dag \gamma^0$.
	
	Now we decorate a spin to each lattice site. The expectation value of the angular momentum of the spin at site $i$ is $\langle\pmb S_i \rangle=\pmb n_i$. We assume that $\pmb n_i$ is a smooth function of the site index $i$. The decorated spins couple to the electrons via the following Hamiltonian 
	\beq\label{couple}
	H_1&=&\sum_{i} f(i)m C_{i}^\dag \pmb \sigma C_{i} \cdot \pmb n_{i} \notag\\
	&=&\sum_{k,q}  m C_{k}^\dag \gamma^z \pmb \sigma C_{q} \cdot \pmb n_{k-q} ,
	\eeq
	with $C^\dag_{i}=(c_{i\up}^\dag,c_{i\dn}^\dag), C_k=\sum_i C_i e^{i\pmb k\cdot \pmb r_i}$,  and $f(i)=1$ if $i \in A$-sublattice and $f(i)=-1$ if $i \in B$-sublattice. In this forms of coupling, the fermions on $A$ and $B$ sublattices feel opposite magnetic momenta, which indicates that at short distance the decorated spins exhibit anti-ferromagnetic correlation.

\begin{figure}[t]
\centering
\includegraphics[width=0.9\linewidth]{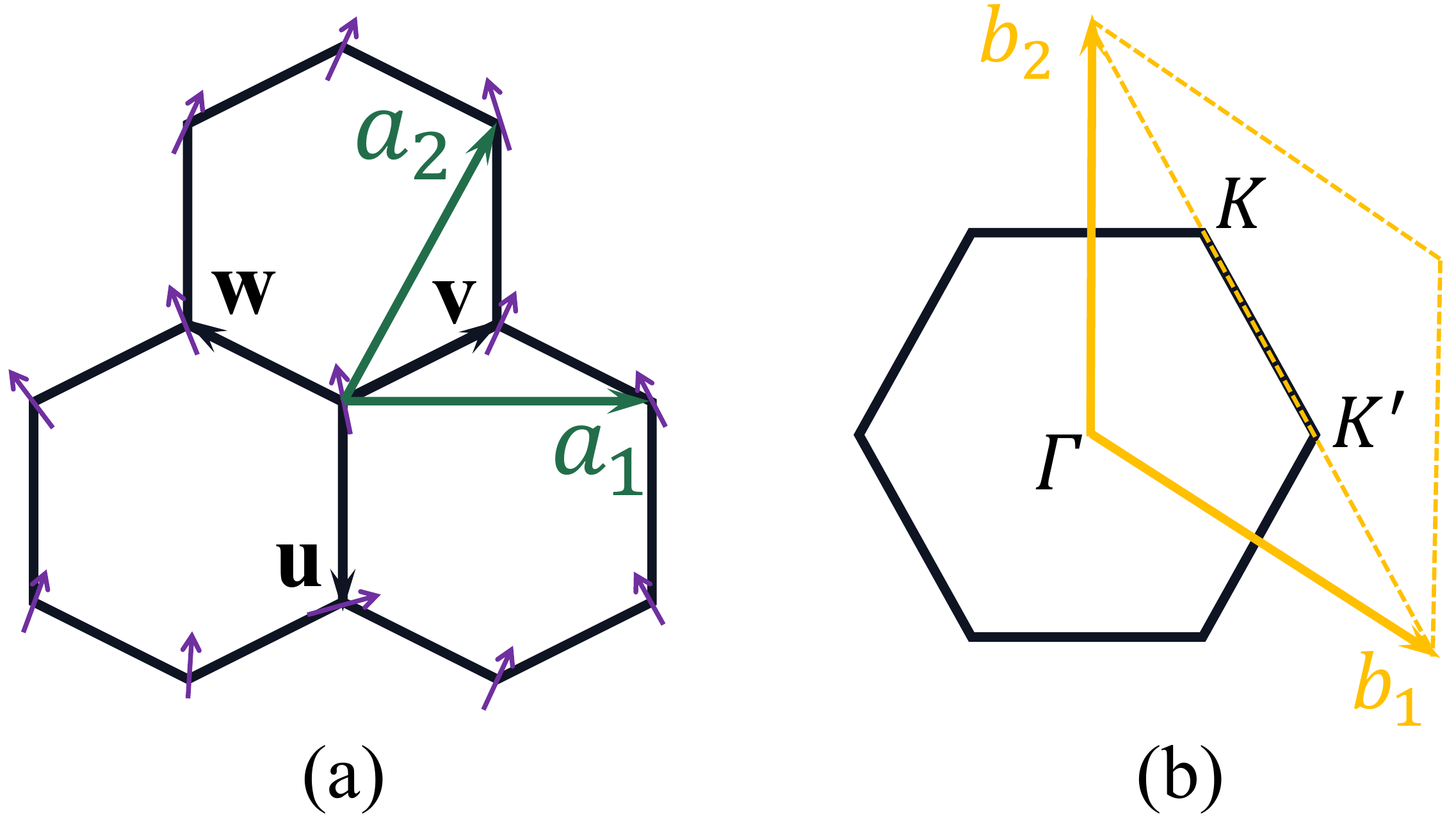}
\caption[4band]{(a) Cartoon picture of the honeycomb lattice. The purple arrows stand for decorated spins on each site. (b) The bases for the reciprocal lattice and the Brillouin zone.
}
\label{fig:honeycomb}
\end{figure}

Projecting onto the low-energy subspace, the effect action reads,  
$$
S_{\rm eff} =\int d^3x \bar\psi (i \gamma^0\partial_0 + i\mathcal V_z\gamma^1\partial_1 + i\gamma^2\partial_2 + m\pmb n\cdot \pmb \sigma + \mathcal L_{\rm int}) \psi,
$$
where $\mathcal L_{\rm int}$ stands for the intercone coupling term. For convenience, we adopt a new set of bases $\xi^\dag (x)=[ \psi^\dag_{KA}(x),\psi^\dag_{KB}(x), \psi^\dag_{K'B}(x), -\psi^\dag_{K'A}(x) ]$ (compared with the original bases, the order of $\psi^\dag_{K'A}$ and $\psi^\dag_{K'B}$ are exchanged, and a minus sign is added to $\psi^\dag_{K'A}$) under which the above action is transformed into
$$
S_{\rm eff} =\int d^3x \bar\xi ( \sum_{\mu=0}^2 i\gamma^\mu\partial_\mu+ \mathcal V_z m\pmb n\cdot \pmb \sigma + \mathcal L'_{\rm int}) \xi.
$$
We firstly ignore $ \mathcal L_{\rm int}'$ such that the two cones are decoupled. The $\mathcal V_z$ matrix in the mass term indicates that the two Dirac cones at $K$,$K'$ have opposite signs of mass. Consequently, the Hopf terms contributed from the two valleys exactly cancel each other and only a dynamic term remains in the end. As shown in the Appendix \ref{app:intercone}, the intercone coupling term $\mathcal L'_{\rm int}$ does not generate any topological term neither.

In summary, the Dirac cones of the $p_z$-orbit electrons on the honeycomb lattice cannot generate nontrivial topological terms for the decorated spins. 
	
\begin{figure}[t]
\centering
\includegraphics[width=0.7\linewidth]{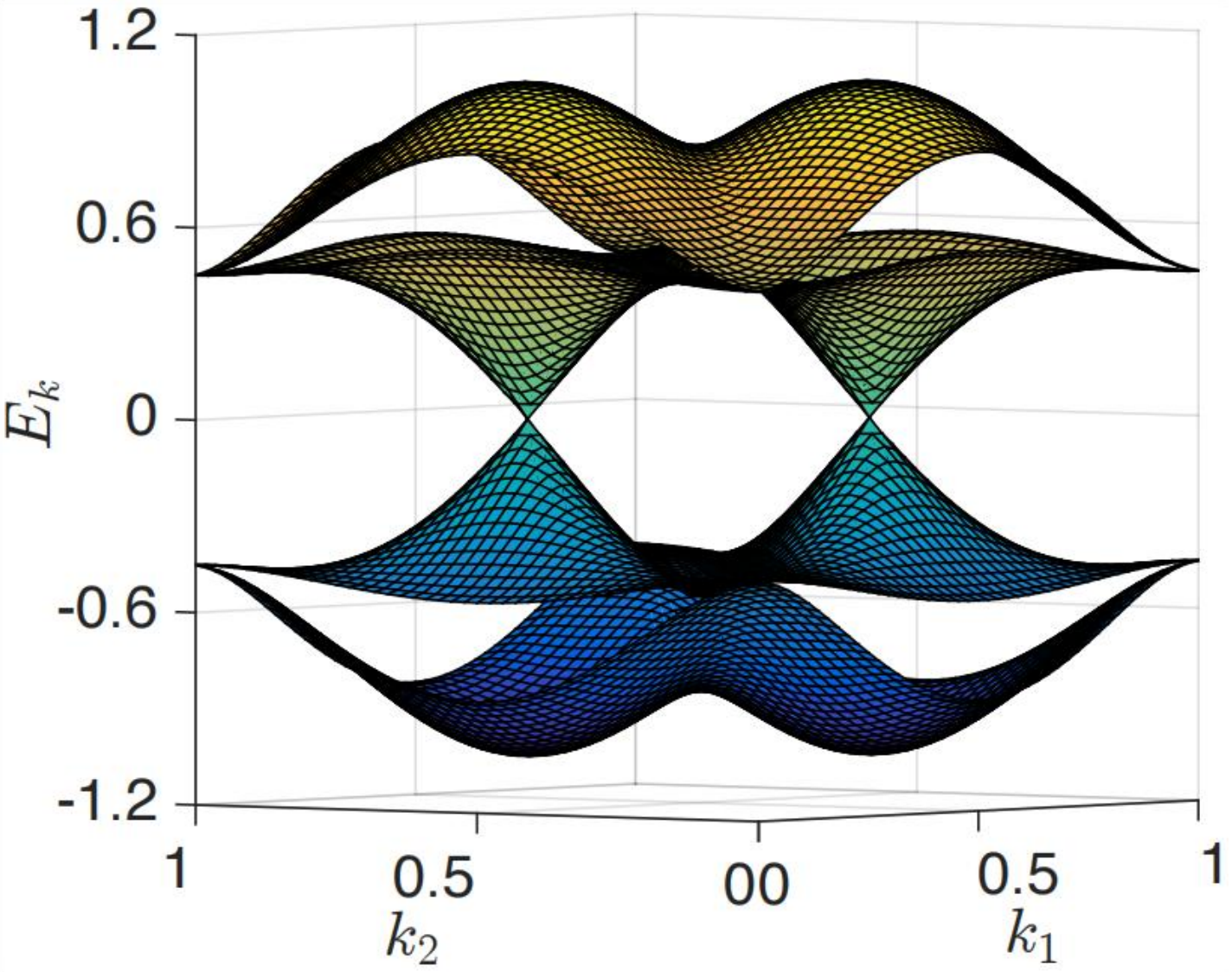}
\caption[4band]{ The Dirac cone band structure of the honeycomb fermion model with $p_x, p_y$ orbitals.}
\label{fig:Dirac}
\end{figure}	
	
	{\bf Hopf term from a four-band model.} Now we consider the $p_x, p_y$ orbitals on the honeycomb lattice\cite{WuCJ1, WuCJ2}. Unlike the $p_z$ orbit which only forms $\pi$-bonds, the $p_x,p_y$ orbits also form $\sigma$-bonds. For convenience, we introduce the eigen bases of the orbital angular momentum operator $l_z$, i.e. $c_{i\pm\sigma}={1\over\sqrt2}(c_{ix\sigma}\pm ic_{iy\sigma})$. We further hide the spin indices and denote $C_{i\pm} = {1\over\sqrt2}(C_{ix}\pm iC_{iy})$. Ignoring the spin-orbit coupling, then the tight binding model with nearest neighbor hopping reads 
	\Beq
	H&=&\sum_{\langle i,j\rangle} \Big( h_{++}^{ij}C_{i+ }^\dag C_{j+} + h_{+-}^{ij} C_{i+}^\dag C_{j-} + h_{-+}^{ij} C_{i-}^\dag C_{j+} + \\
	&&\ \ h_{--}^{ij} C_{i-}^\dag C_{j-} + {\rm h.c.}\Big),
	\Eeq
	where $h_{\pm\pm}^{ij}$ are bond-dependent hopping constants with the relation $h_{+-}^{ij}=(h_{+-}^{ji})^*=h_{-+}^{ji}$. We label the three bonds linking A-sublattice and B-sublattice as $\pmb u=(0,-{\sqrt3\over3})={1\over3}\pmb a_1-{2\over3}\pmb a_2, \pmb v=({1\over2}, {\sqrt3\over6})={1\over3}\pmb a_1+{1\over3}\pmb a_2, \pmb w=(-{1\over2}, {\sqrt3\over6})=-{2\over3}\pmb a_1 + {1\over3}\pmb a_2$ [as shown in Fig.\ref{fig:honeycomb}(a)], then 
	\Beq
	&&h_{++}^{ij}=h_{--}^{ij} = {1\over2}(V_\pi+V_\sigma),\ \  h_{+-}^{ij=u}= {1\over2}(V_\pi-V_\sigma), \\
	&&h_{+-}^{ij=v}= {1\over2}(V_\pi-V_\sigma)e^{i2\pi/3},\ \ \ h_{+-}^{ij=w}= {1\over2}(V_\pi-V_\sigma)e^{i4\pi/3},
	\Eeq 
	where $V_\sigma,V_\pi$ are the hopping integral of the $\sigma,\pi$ bonds respectively.  Some materials such as the Ge or As layer on the SiC substrate are approximately described by the above model\cite{WuCJ2}.

The above model has four energy bands, the intermediate two bands touch each other at $K, K'$ and form Dirac cones respectively (see Fig.\ref{fig:Dirac}). The Hamiltonian can be written in momentum space as 
	\beq\label{bareH}
	H = \sum_k   C_k^\dag \mathcal H_k  C_k 
	\eeq
	where $  C_k^\dag=(C_{kA+}^\dag, C_{kA-}^\dag, C_{kB+}^\dag, C_{kB-}^\dag), \mathcal H_k=\Bmat 0&h_{k}\\h_{k}^\dag &0 \Emat$ with $h_k$ a two-by-two matrix whose entries are given by
	\Beq
	&&(h_k)_{11} =(h_k)_{22} = {1\over2}(V_\pi+V_\sigma) [1+2 \cos{k_x\over2} e^{i{\sqrt3\over2}k_y}],  \\
	&&(h_k)_{12} =  {1\over2}(V_\pi-V_\sigma) [1+2 \cos ({k_x\over2} + {2\pi\over3}) e^{i{\sqrt3\over2}k_y}], \\
	&&(h_k)_{21} =  {1\over2}(V_\pi-V_\sigma) [1+2 \cos ({k_x\over2} - {2\pi\over3}) e^{i{\sqrt3\over2}k_y}].
	\Eeq
	
	At the $K$ and $K'$ points, $h_k$ reduces to $h_K = \Bmat 0 & {3\over2}(V_\pi-V_\sigma)\\0&0\Emat$ and $h_{K'} = \Bmat 0 & 0\\ {3\over2}(V_\pi-V_\sigma)&0\Emat$, respectively. It is obvious that the energy eigenvalues of $\mathcal H_k$ at $K, K'$ are ${3\over2}(V_\pi-V_\sigma), 0,0,-{3\over2}(V_\pi-V_\sigma)$, where the zero energies modes give rise to two Dirac cones which connect the second and the third bands. 
	
	It can also be seen that the zero-energy eigenspace at $K$ is spanned by $C^\dag_{K,A-},C^\dag_{K,B+}$, while at $K'$ the zero-energy eigenstates are $C^\dag_{K'A+}, C^\dag_{K'B-}$. Since the low-energy physics is determined by the quasiparticles in the vicinity of the cones in the two intermediate bands, it will be convenient to introduce the following bases for the low-energy subspaces
	\Beq
	\Psi_{k}^\dag&=&\Big( \psi^\dag_{K+k,B+\up}, \psi^\dag_{K+k,B+\dn}, -\psi^\dag_{K+k,A-\up}, -\psi^\dag_{K+k,A-\dn},\\
	&&\psi^\dag_{K'+k,A+\up}, \psi^\dag_{K'+k,A+\dn}, \psi^\dag_{K'+k,B-\up}, \psi^\dag_{K'+k,B-\dn} \Big).
	\Eeq 
	
	We further denote the matrices $\mathcal V_{x,y,z}, \gamma^{x,y,z}$ to act on the valley index ($K$ and $K'$) and the band index (correspond to the second and the third bands in the original band structure which hosts the Dirac cones), respectively.  Projecting the original Hamiltonian (\ref{bareH}) onto the low-energy subspace, we obtain the  effective $k.p$ Hamiltonian for small $k$,
	\Beq
	H_{\rm eff}  &=& \sum_{k} \Psi_{k}^\dag \mathscr H_{k}\Psi_{k},\\
	\mathscr H_{k}&=&v ( k_x\gamma^x+ k_y\gamma^y).
	\Eeq
	with $v={\sqrt3\over4}(V_\pi-V_\sigma)<0$.
	
	Again, we decorate a spin to each site and couple them to the fermions. But if the coupling takes the same form of Eq.(\ref{couple}), namely, electrons on $A$ and $B$ sublattices fell opposite magnetic momentum of the decorated spin, then the topological terms contributed from the two Dirac cones will cancel each other as happened to the graphene. 
	
However, if the direction of the spin momentum $\pmb n(x)$ is sensitive to the orbital angular momentum $l_z=\pm1$ such that the electrons with $l_z=1$ and $l_z=-1$ feel the opposite magnetic momenta from the decorated spin, namely, if the coupling takes the following form (we have put back the spin index)
\beq
H_1 &=& \sum_i m \big(C_{i  +}^\dag\pmb \sigma C_{i +} \cdot \pmb n_{i }- C_{i  -}^\dag\pmb \sigma C_{i -} \cdot \pmb n_{i }\big)\nonumber\\
	&=&\sum_{k,p} m  (C_{k A+}^\dag  \pmb \sigma  C_{pA+} -C_{k A-}^\dag  \pmb \sigma  C_{pA-} \nonumber\\
	&&\ \ \ \ \ \ + C_{k B+}^\dag  \pmb \sigma  C_{pB+} - C_{k B-}^\dag  \pmb \sigma  C_{pB-}) \cdot \pmb n_{k-p}. \label{couple2}
\eeq
Then the Hopf terms contributed from the two cones may have the same sign and there will be a nonzero topological term for the $\pmb n_i$ field after integrating out the fermions.

To verify the above conjecture, we will explicitly integrate out the fermions. Projecting the coupling term (\ref{couple2}) onto the low-energy subspaces, we obtain
\Beq
PH_{\rm 1}P^{-1} = m\sum_{k,p} \pmb n_{k-p}  \cdot\Psi_k^\dag  \pmb\sigma \mathcal \gamma^z \Psi_p,
\Eeq 
where $P$ stands for the projection operator onto the subspace spanned by the bases $\Psi^\dag_{k=0}$.
Noticing that the intercone scattering process vanishes automatically, therefore $\pmb n$ field cannot couple the $\Psi_{K+k}$ fermions with $\Psi_{K'+p}$ fermions. 
	
	Rescaling the fermi velocity $|v|=1$ and introducing the matrices $\gamma^0=-\gamma^z, \gamma^1= -\gamma^0 \gamma^x, \gamma^2=-\gamma^0 \gamma^y$ with $\{\gamma^\mu, \gamma^\nu\}=g^{\mu\nu}$, then in the continuum limit the effective action of the coupled system reads (for details see the Appendix),
	\beq
	 S_{\rm eff} = \int d^3x \bar\Psi(x) \big[\sum_{\mu=0}^2i\partial_\mu\gamma^\mu + m\pmb n(x)\cdot\pmb \sigma \big] \Psi(x),
	\eeq
	where $\bar\Psi=\Psi^\dag \gamma^0$.
	In the above effective action, the two valleys couples to the vector field $\pmb n(x)$ in the same way. 
 
Therefore, from the previous discussion,  we finally obtain the Hopf model (\ref{O3nlsm}) with $\theta=2\pi$ after integrating out the fermions,
\beq\label{hopf2}
S_{\rm hopf} = 2\pi\ {\rm sgn}(m) H(\pmb n),
\eeq
where $H(\pmb n)$ is defined in Eq.(\ref{Hopfind}).

{\bf $\theta=\pi$ term from fermions at the critical point}. In the above discussion, the spins are coupled to a pair of Dirac cones. Under fine tuning, it is possible to gap out one of the cones (e.g. by breaking some symmetry) and keep the rest cone massless. Another possibility is that at the critical point, for instance from a trivial band insulator to a Chern insulator with Chern number $C=1$, there will be a single Dirac cone in the first Brillouin zone. In this case, if we couple the spins to the 
fermions, then a Hopf term with $\theta=\pi$ may be obtained, in which case a skyrmion is interpolated by a fermion doublet.   
 
However, if the ground state does not spontaneous break the $SO(3)$ symmetry, then the fermions cannot dynamically obtain a mass\cite{IJMPA1990}. Therefore, it is likely that the ground state belongs to a gapless phase\cite{IJMPA1990, XuLudwig}. 

In contrast, when $\theta=2\pi$ and if the ground state does not break the $SO(3)$ symmetry, a skyrmion excitation traps two fermions in its core.  The two fermions form a spin triplet owing to the spin-Hall effect (see Sec.\ref{3bb}) which costs a finite pair-breaking energy recalling that the ground state is spin-singlet. Hence, the intrinsic excitations are bosonic and gapped.
	
\section{Consequence of the Hopf term}\label{sec:spt}
	
\subsection{The Hopf model and the principal chiral NLSM}
In the previous discussion, the Hopf term was written in terms of $a_\mu$ which is nonlocal in  forms of $\pmb n$. It was shown that  the Hopf term is local in forms of the spinor field $z$\cite{WuZee}, with $a_\mu = -i z^\dag \partial_{\mu} z$. Now we introduce an $SU(2)$ element 
\beq\label{gEuler}
g(x)=\exp\{-i{\sigma^z\over2}\varphi\} \exp\{-i{\sigma^y\over2}\theta\} \exp\{-i{\sigma^z\over2}\omega\}
\eeq  
such that 
\beq\label{zg}
z=g\Bmat1\\0\Emat
\eeq
and $\pmb n=z^\dag \pmb \sigma z =(\sin\theta\cos\varphi, \sin\theta\sin\varphi, \cos\theta)^T$. The correspondence from $g$ (or $z$) to $\pmb n$ is many to one because the angle $\omega$ does not appear in $\pmb n$.  Introducing the Berry connection $A_\mu=g^{-1} \partial_\mu g = \sum_m A_\mu^m\sigma^m$,  it is easily checked that $a_\mu$ in Eq.(\ref{Hopfind}) is the $z$-component of $A_\mu$,  namely $A_\mu^z= a_\mu$.

Furthermore, it can be verified that $\varepsilon^{\mu\nu\lambda}A^x_\mu\partial_\nu A^x_\lambda = \varepsilon^{\mu\nu\lambda}A^y_\mu\partial_\nu A^y_\lambda = \varepsilon^{\mu\nu\lambda}A^z_\mu\partial_\nu A^z_\lambda$, so the Hopf term (\ref{hopf2}) can also be written as 

\beq\label{hopf=pc}
 S_{\rm hopf} &=&- {2\pi\over 24\pi^2} \int d^3x \varepsilon^{\mu\nu\lambda} {\rm Tr} (A_\mu \partial_\nu A_\lambda)\notag\\
&=&- {2\pi \over 24\pi^2} \int  {\rm Tr} (g^{-1} dg)^3.
\eeq  

Similarly, the dynamical term of the $SO(3)$ NLSM can also be written informs of $SU(2)$ group elements (see App.\ref{app:eom}), 
\beq\label{SO3dyn}
S_{\rm SO(3) dyn} &=& \int d^{3}x {1\over\lambda} \Big((\partial_t\pmb n)^2 - (\nabla \pmb n)^2\Big)\notag\\
 &=&\int d^3 x\frac{1}{\lambda} {\rm Tr} (\partial^\mu g\partial_\mu g^{-1} + g\sigma^z\partial^\mu g^{-1}g\sigma^z\partial_\mu g^{-1}),\notag\\
\eeq  

It is necessary to clarify the symmetry group and the way it acts on the variables. It seems that the Hopf model (\ref{O3nlsm}) has $SO(3)$ symmetry, but under the condition $z^\dag z=1$ the field $\pmb n(\pmb r)$ actually describes $S=1/2$ spins, so the symmetry group is better identified as $SU(2)$. Supposing $h\in SU(2)$ is a symmetry operation, then it acts on $\pmb n$ in the following way,
\Beq
h \hat n h^{-1} = \pmb n\cdot h\pmb\sigma h^{-1} = \sum_{i,j} n_i D^{\rm(vec)}(h)_{ji} \sigma_j, 
\Eeq
where $D^{\rm(vec)}(h)$ is the vector representation of $h$ and
\[
\hat n = \pmb n\cdot\pmb\sigma = 2z z^\dag -I =g \sigma^z g^{-1}.
\]
 So under the symmetry operation $h$, $\pmb n$ varies as a vector $n_j\mapsto \sum_{i} D^{\rm(vec)}(h)_{ji}n_i $  and  
 $g$ varies as 
\beq 
 g\mapsto hg.
\eeq 
Namely, $h$ acts on $g$ by {\it left multiplication}. We call the group formed by these symmetry operations as $SU(2)_L$.

The action (\ref{SO3dyn}) together with Eq.(\ref{hopf=pc}) is closely related to the  O(4) NLSM with theta term \cite{you2015bridging,you2018bosonic,xu2013wave} or the
$SU(2)$ principal chiral NLSM\cite{XuLudwig, LiuWen} with $\mathcal K=-1$,
\beq\label{SU2pc}
S_{\rm SU(2)pc} = \int d^3 x \frac{1}{\lambda} {\rm Tr} (\partial^\mu g\partial_\mu g^{-1}) - {2\pi \over 24\pi^2} \int  {\rm Tr} (g^{-1} dg)^3. \notag\\
\eeq
Actually, the Hopf model and the SU(2) principal chiral NLSM model share the same topological term but differ by their dynamical terms. 

The different dynamical terms result in different symmetry groups. We have shown that the Hopf model has $SU(2)_L$ symmetry. But the $SU(2)$ principal chiral NLSM is invariant under both left multiplication $g\mapsto hg$ and right multiplication $g\mapsto gh^{-1}$, thus the symmetry group of Eq.(\ref{SU2pc}) is $SU(2)_L\times SU(2)_R/Z_2\simeq SO(4)$.

However, if the coupling constant $\lambda$ of the dynamic term is initially not very small, it may flow to infinity $\lambda\to \infty$ under renormalization group (RG). Consequently the system falls in a topological phase where  the ground state and the low energy physics are dominated by the topological $\theta$-terms. In this limit, the Hopf model and the $SU(2)$ principal chiral NLSM have similar physical properties.
In the following we will show that in the strong coupling limit the ground state of the Hopf model 
belongs to a SPT phase.

\subsection
{The root SU(2) SPT phase} $\label{3bb}$
As pointed out in Ref.\onlinecite{WilczekZee}, in the presence of a Hopf term $S_{\rm hopf}=\theta H(\pmb n)$, a skyrmion-type solition has angular  momentum $\theta/2\pi$ and hence has statistical angle $\theta$. Since the Hopf term (\ref{hopf2}) derived from the lattice model has $\theta=2\pi$, so the skyrmions obey bosonic statistics under braiding. This suggests that the resultant ground state is possibly a SPT state.

In Ref.\onlinecite{LiuWen}, it was shown that in the strong coupling limit the ground state of $SU(2)$ principal chiral NLSM is both a $SU(2)_L$ SPT phase and an $SU(2)_R$ SPT phase. Notice that the second part of the dynamical term (\ref{SO3dyn}) in the Hopf model breaks the $SU(2)_R$ symmetry but preserves the $SU(2)_L$ symmetry. Therefore, the strong coupling phase of the Hopf model is essentially an $SU(2)_L$ SPT phase. In later discussion, we will eliminate the subscript $L$ and will simply call the symmetry group $SU(2)_L$ as $SU(2)$ when it does not cause confusion.

The $SU(2)$ SPT phases have $\mathbb Z$ classification where each value of $\mathcal K\in\mathbb Z$ stands for a distinct SPT phase. Since the $\theta=2\pi$ Hopf model corresponds to the $\mathcal K=-1$ phase, thus it is the root phase of $SU(2)$ SPT phases because $\mathcal K=-1$ generates the $\mathbb Z$ classification.

The $SU(2)$ SPT phases are characterized by their gapless edge excitations (if the symmetry is unbroken) and the nontrivial spin Hall effect. The spin Hall conductance according to the probe field $\mathcal A^z$ is quantized to\cite{LiuWen} 
$$
\sigma_{xy}^z = {\mathcal K\over 4\pi} = 2\mathcal K{S^2\over 2\pi},
$$
which is an even integer times ${S^2\over 2\pi}$ with $S=1/2$. The spin Hall effect is not a surprise since the Hopf term is in the same form with the Chern-Simons term for gauge fields. The Chern-Simons term is the low energy effective theory of Hall effects and its edge excitation spectrum is chiral.  However, as an SPT state, the energy spectrum should be nonchiral. It seems to be a paradox that spin Hall effect is chiral but the energy spectrum of the edge is nonchiral. 

Noticing that the bulk topological $\theta$ term provides a Wess-Zumino-Witten (WZW) term of the edge, the low energy theory of the edge can be described by 
\Beq
S_{\rm edge} &=&\int dxdt \Big({1\over\lambda} {\rm Tr} (\partial^\mu g\partial_\mu g^{-1}) + \mathcal L_{\rm int}\Big) \\
&&- {2\pi \over 24\pi^2} \int  {\rm Tr} (g^{-1} dg)^3,
\Eeq
where the $\mathcal L_{\rm int}$ term breaks the $SU(2)_R$ symmetry and results in a dynamical term of the $SO(3)$  NLSM\cite{WZW}. It was shown that the $SO(3)$ NLSM plus the $\mathcal K=-1$ WZW term can be identified with the $(1+1)$D $SO(3)$ NLSM plus a $\theta=\pi$ topological term, in which the energy spectrum is gapless and the critical theory falls in the same class of the $\mathcal K=-1$ WZW model. Therefore, in the low-energy limit, we can set $\mathcal L_{\rm int}=0$ and $\lambda={{8\pi}}$ at the fixed point.

From the non-Abelian bosonization theory\cite{Witten_Bosonization}, at the fixed point the WZW model decouples into gapless left mover $J_+ =i {1\over 2\pi}\partial_+g g^{-1}$ and right mover $J_-=-i{1\over2\pi} g^{-1}\partial_-g$ with $\partial_\pm = {1\over\sqrt2}(\partial_t \mp \partial_x)$. The left mover $J_+$ carries $SU(2)_L$ charge (since it is covariant under the action $g\mapsto  hg$) and the right mover $J_-$ is a $SU(2)_L$ singlet (since it is invariant under the action $g\mapsto hg$),
and they satisfy the equation of motion $\partial_{+}J_{-}=0$ and $\partial_{-} J_+=0$  as long as the $SU(2)_L$ symmetry is unbroken. Hence, if the boundary of the Hopf model preserves all the symmetries, then the edge theory of the Hopf model will flow to the fixed point of the WZW model whose gapless excitations are nonchiral in energy but chiral in symmetry. 
 
The gapless spectrum and nonzero spin Hall effect reflects that the edge theory is anomalous and cannot be realized in pure 1D given that the symmetry acts in a local manner.

On the other hand, if the band structure of the fermions contains more than two Dirac cones, then the spin system coupled to the fermions may acquire a Hopf term with $\theta = 2\pi \mathcal K$ with $|\mathcal K|\geq 2$. The resultant ground state belongs to the $SU(2)$ SPT phase in the $\mathcal K$th class. Alternatively, the $\mathcal K$th class $SU(2)$ SPT phase can also be obtained by stacking $\mathcal K$ layers of root phases which are weakly coupled to each other.

\subsection{ SO(3) SPT for integer spins}

In the following, we discuss the case where there are still two Dirac cones but the decorated spin is larger than one-half $S>1/2$. In this case, the derivation of Eq.(\ref{hopf2}) remains valid besides that we should take the following replacement $z\to \sqrt{2S}z, a_\mu\to 2S a_\mu, m\to 2Sm$, and 
\beq\label{spinS}
S_{\rm hopf} (S)=  2\pi (4S^2) \ {\rm sgn}(m) H(\pmb n).
\eeq
This corresponds to a $SU(2)$ SPT phase with $\mathcal K=-4S^2$. The spin Hall conductance according to the probe field $\mathcal A^z$ is 
$$
\sigma_{xy}^z={\mathcal K\over 4\pi}=-2S^2{1\over2\pi}.
$$
	
If $S$ is an integer, then it is natural to identify the symmetry group as $SO(3)$. Introducing the matrix $\tilde A_\mu = h^{-1}\partial_\mu h=\sum_m \tilde A_\mu^m L^m$, where $h\in SO(3)$ and  $L^{x,y,z}$ are three generators of $SO(3)$. Then we have $\tilde A_\mu^z = 2a_\mu$, and the topological term (\ref{spinS}) can be written as
\Beq
	 S_{\rm hopf} &=& -2\pi (4S^2)  {1\over 4\pi^2} \int d^3x \epsilon^{\mu\nu\lambda} a_\mu\partial_\nu a_\lambda \\
	&=&   -2\pi (4S^2) {1\over 16\pi^2 }\int d^3x \epsilon^{\mu\nu\lambda}  \tilde A^z_\mu\partial_\nu \tilde A ^z_\lambda \\
	&=&    -2\pi (4S^2) {1\over 2\times 48\pi^2}\int d^3x \epsilon^{\mu\nu\lambda}  {\rm Tr}[\tilde A_\mu\partial_\nu \tilde A_\lambda]\\
	&=&   -2\pi (4S^2) {1\over 2\times 48\pi^2}\int  {\rm Tr} (h^{-1}d h)^3
	\Eeq

This is the $SO(3)$  principle chiral NLSM with $\mathcal K=-4S^2$,  which corresponds to a $SO(3)$ SPT phase (which requires $\mathcal K$ is multiple of 4) with the spin Hall conductance 
$$
\sigma_{xy}^z = {\mathcal K\over 4\pi} = -2S^2{1\over 2\pi},
$$
which is always quantized to an even integer in unit of ${1\over 2\pi}$. Compared with the  previous discussion, the spin Hall conductance remains the same if the symmetry group is interpreted as $SU(2)$. 
As expected, for integer spin system there is no essential difference to consider the symmetry group as $SO(3)$ or $SU(2)$.

\section{Hopf term from Massive Dirac fermions}\label{sec:gapped}

\subsection{With a constant mass $m_0$}\label{sec:m0}	
In the above discussion, we have assumed that the Dirac fermions are massless before coupling to the spins.In the following we add a constant mass term such that the Dirac fermions open a gap. If the Chern number is zero, then the effective Lagrangian reads 
$$
S_{\rm eff} =\int d^3x\ \bar\Psi ( \sum_{\mu=0}^2 i\gamma^\mu\partial_\mu + m_0 \mathcal V_z )
\Psi;
$$
if the Chern number is $\pm1$, then the effective Lagrangian is
$$
S_{\rm eff} =\int d^3x\ \bar\Psi ( \sum_{\mu=0}^2 i\gamma^\mu\partial_\mu + m_0)
\Psi.
$$
The mass $m_0$ may be generated by symmetry  breaking perturbations or result from spin-orbit coupling (SOC).  

\begin{figure}[t]
\centering
\includegraphics[width=0.9\linewidth]{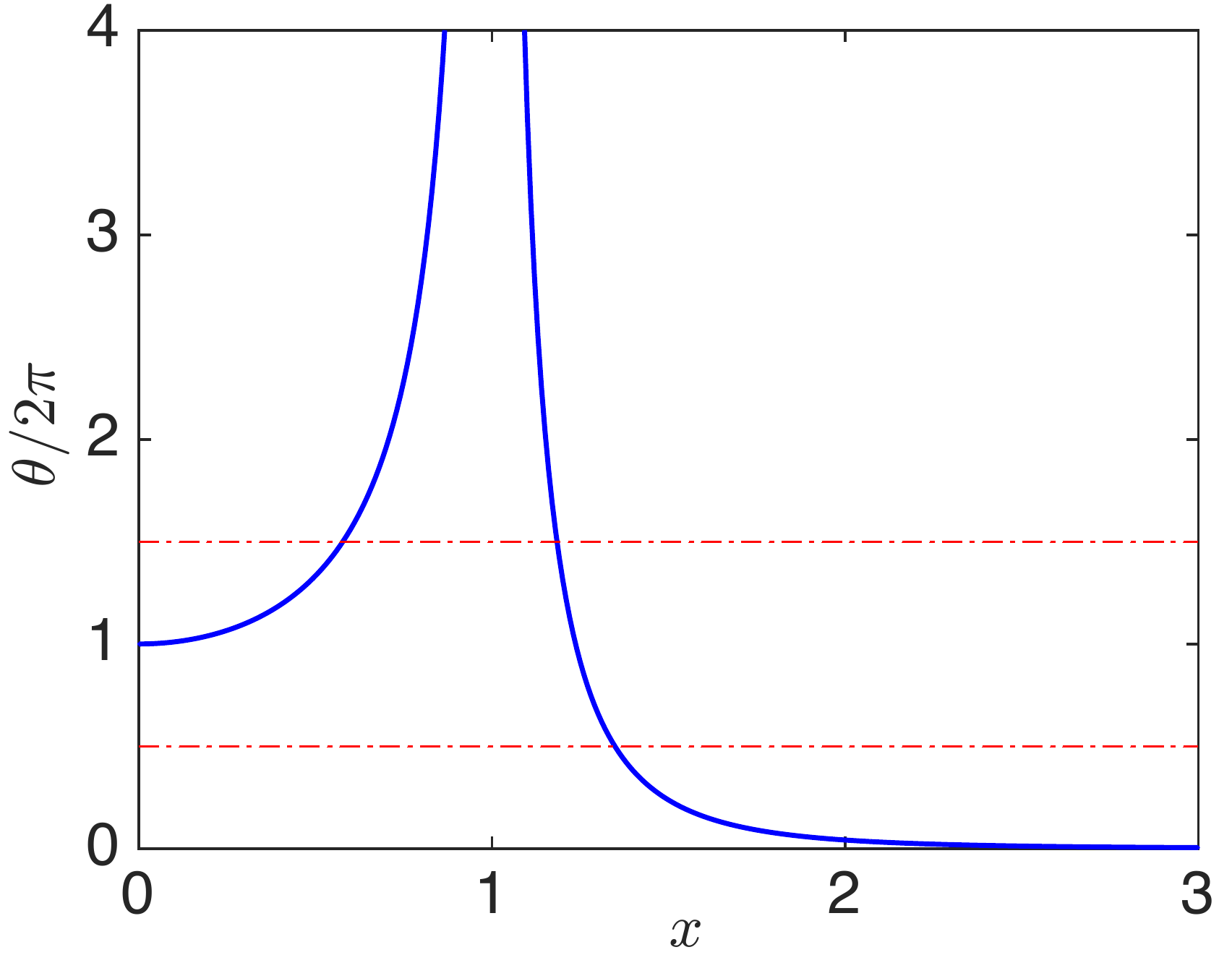}
\caption[4band]{The value of $\theta(x)$ as a function of $x={m_0\over m}$ with $\theta(-x)=\theta(x)$. Here we have summed over the first $M=100000$ terms in the series (\ref{theta_x}). The function $\theta(x)$ diverges at $|x|=1$ with increasing $M$. Between the two dashed red lines, $\theta$ flows to $2\pi$ under RG. Blow the lower dished red line, $\theta$ flows to 0. } 
\label{fig:theta}
\end{figure}	
	
Firstly we assume that the $m_0$ mass term does not break the $SU(2)$ symmetry. When coupled to the decorated spins, we should add $m\bar\Psi \hat n \Psi$ to the Lagrangian. Due to the $m_0\bar\Psi \Psi$ term, when deriving the effective theory one should expand the inverse of the operator 
$$
D\bar D =\partial^2 +m_0^2+m^2 +2m_0m\hat n + im\gamma^\mu\partial_\mu \hat n
$$
in polynomials of ${1\over \partial^2 + m_0^2+m^2}$ (see Appendix \ref{app:m0} for details). Except for the case where $m_0=0$, in the above expansion there are infinite terms contributing to the  topological term. The sum of all these terms determines the value of $\theta$. Denoting $x={m_0\over m}$, then  
\beq
&&\theta(x) = -2{\rm sgn}(m)\cdot32\pi^2 \sum_{N=1}^{\infty} f_{2N+1}\label{theta_x},\\
&&f_k(x)=- {x^{k-3}\over   \big(\sqrt{1+x^2}\big)^{2k-1}}\eta(k), \notag
\eeq
with $\eta(k)= \frac{2^{k-4}(k-1)}{4k\pi} \frac{(2k-3)!!}{(2k-2)!!} $. Interestingly, the value of $\theta$ is independent on the sign of $m_0$. The numerical estimation of $\theta(x)$ in  Eq.(\ref{theta_x}) is shown in Fig. \ref{fig:theta}.

Notice that the two mass terms $m_0\bar\Psi\Psi$ and $m\bar\Psi\hat n\Psi$ commute with each other. When the two mass terms are close in magnitudes, namely when $|{m_0\over m}|\to 1$, then the total energy gap will approach to zero. Consequently, the series (\ref{theta_x}) diverge as $|{m_0\over m}|\to 1$.  
Away from this singular point $|{m_0\over m}|=1$, the series converges well and the sum is a continuous function as $|{m_0\over m}|<1$ or $|{m_0\over m}|>1$. For reason that will be given later, the special values ${\theta= 2N \pi}$ are of importance and each defines a class. The values of $\theta$ locating in the vicinity of $2N\pi$ with $|\theta-2N\pi|<\pi$ are considered to belong to the class associate with $\theta=2N\pi$. For instance, in the region $|{m_0\over m}|<0.57$, $|\theta({m_0\over m}) - \theta(0)|<{\pi}$ with $\theta(0)=2\pi$. Therefore the region $|{m_0\over m}|<0.57$ is considered as belonging to the same class with $m_0=0, \theta=2\pi$. When $0.57<|{m_0\over m}|<1$, the value of $|\theta|$ increases rapidly and eventually blows up.

On the other side, when $|{m_0\over m}|$ exceeds 1, $\theta({m_0\over m})$ 
decreases rapidly from $\infty$ to 0 with increasing $|{m_0\over m}|$. In the region $|{m_0\over m}|>1.35$,  $|\theta({m_0\over m}) - \theta(\infty)| < {\pi}$ where $\theta(\infty)=0$. Therefore the region $|{m_0\over m}|>1.35$ is considered as belonging to the trivial class with $m_0=\infty, \theta=0$.

The criterion for the equivalence classes for the values of $\theta$ is bases on RG. Here we assume that in the strong coupling phase (where $\lambda$ is big such that the dynamic term is unimportant) the RG flow of the Hopf model (\ref{O3nlsm}) is similar to that of the $SU(2)$ principal chiral NLSM \cite{XuLudwig}. This assumption is reasonable because the two models have the same topological term and in the strong coupling phase the topological term dominates the low energy physics.
 
With the vanishing of the dynamic term, the value of $\theta$ flows to a nearby fixed point. According to Ref.\onlinecite{XuLudwig},  the fixed points include the unstable ones $\theta =(2N+1)\pi$ and the stable ones $\theta=2N\pi$ with $N\in\mathbb Z$. 

From the above  expression (\ref{theta_x}) of $\theta$, it can be inferred that if $|{m_0\over m}|<0.57$, then $|\theta-2\pi|<\pi$, it will flow to a stable fixed point $2\pi$. Therefore, a small constant mass with $|{m_0\over m}|<0.57$ has the same effect with $m_0=0$ in the sense of RG. For this reason, the SPT phase corresponding to $\theta=2\pi$ is robust against small perturbations to the fermions. Similarly, a large constant mass results in a vanishing topological term for the spins because when $|{m_0\over m}|>1.35$,  the absolute value of $\theta$ is less then $\pi$ and will flow to 0 under RG.

\subsection{The effect of spin-orbit coupling}	

Now we briefly discuss the effect of SOC in the fermions. SOC has two direct consequences: (I) the Dirac cones at the $K ,K'$ points are gapped out; (II) the $SU(2)$ spin rotation symmetry is gone and the resultant symmetry group is a discrete group. For simplicity, we only consider the term descendant from $\lambda_{\rm so}\pmb S\cdot\pmb L$. 
 
In the following, we firstly illustrate that the topological term is still present when SOC is considered. Then we show that even if the symmetry group is no longer $SU(2)$, the resultant ground state is still a SPT state which is protected by a discrete group. 

We treat the $\lambda_{\rm so}$ term as perturbation, and project it onto the two intermediate bands near the fermi level. It turns out that when $\lambda_{\rm so}$ is not big then the $K$ valley is still dominated by the $p_x+ip_y$ orbit, and the $K'$ valley is dominated by the $p_x-ip_y$ orbit. In the presence of $\lambda_{\rm so}$, one should expand the inverse of the operator 
$$
D\bar D=\partial^{2}+m^{2}+\lambda_{\rm so}^{2}+\lambda_{\rm so}m\{\sigma^{z},\hat n\}+im\gamma^{\mu}\partial_{\mu}\hat n
$$ 
in polynomials of $\frac{1}{\partial^{2}+m^{2}+\lambda_{\rm so}^{2}}$.  As derived in the Appendix \ref{app:SOC}, the topological term resulting from SOC is qualitatively the same as the one with a constant $m_0$ discussed in section \ref{sec:m0}. The raw value of $\theta$ is generally not quantized but it can flow to a nearby quantized value under RG. Specially, when $\lambda_{\rm so}$ is small (i.e. ${\lambda_{\rm so}\over m}<0.83$), 
then $\theta$ will flow to $2\pi$.

As mentioned, SOC reduces the symmetry group to a discrete one. The honeycomb lattice has a six fold rotation $C_6$ symmetry, when considering SOC this $C_6$ symmetry is still present given that the symmetry operation is a combination of six-fold lattice rotation and the corresponding spin rotation. For simplicity, we will consider the two fold rotation $C_2=(C_6)^3$, which form a $Z_2$ group. Furthermore, the classification of SPT phases protected by a point group symmetry can be obtained by treating the point group as an on-site symmetry group\cite{Else}. The physical properties of SPT phases for a point group symmetry is also parallel to those of the SPT phases for the corresponding on-site  symmetry\cite{ZhangNing21, Else, PhysRevB.99.205120} .Therefore, in the following we just treat the above $Z_2$ group as an on-site symmetry group.

Supposing that we firstly turn off SOC such that the effective model is the Hopf model with $\mathcal K=-1$. The corresponding SPT phase has a nontrivial spin quantum Hall effect with Hall conductance $\sigma_{xy}^z= 2\mathcal K{S^2\over 2\pi}$. Consequently, a probe field with spin flux $2\pi\over S$\footnote{Here we have considered the flux quanta as ${2\pi\over S}$ instead of $2\pi$. This is equivalent to regard the fundamental spin $S=1/2$ as the unit spin. In this sense, the two-fold rotation $C_2$ still generates a $Z_2$ group. Otherwise, if we treat $S=1/2$ as half of the unit spin, then $(C_2)^2=-1$ so the group structure becomes $Z_4$.} (this is equivalent to gauging the $SU(2)$ symmetry) is `associated' a spin quantum number $\sigma_{xy}^z {2\pi\over S}=2\mathcal K S$ owing to the spin Hall effect. Since the symmetry group would be reduced to $Z_2$ by SOC, it is interesting to consider a nontrivial $Z_2$ flux, namely ${\pi\over S}$. Obviously, a $Z_2$ flux ${\pi\over S}$ is `associated' spin $-S$. Furthermore, since the symmetry group is $SU(2)$, the $Z_2$ flux can be `attached' with spin quantum number $nS$ with $n\in\mathbb Z$. Therefore, considering the `associated' spin $\mathcal KS$ and the `attached'  spin $nS$, the statistical angle by braiding two $Z_2$ fluxes is given by\cite{WangNormandLiu19}
\Beq
\phi &=& {\pi\over S}\cdot {\mathcal KS\over2} + {\pi\over S}\cdot nS \\
& =& \Big({\mathcal K\over2} + n\Big)\pi.
\Eeq
The first term comes from the `associated' spin $\mathcal KS$ resulting from the spin  Hall effect. The factor ${1\over2}$ is owing to the fact that only one of the two phases, namely the Berry phase by moving the `associated' spin by a semicircle around the flux and the Berry phase by moving the flux by a semicircle around the `associated' spin, need to be counted \cite{WenBook04}. The second term comes from the `attached' spin which is not related to the spin Hall effect. So the factor $1\over2$ is absent. The above expression of the statistical angle $\phi$ indicates that when $\mathcal K$ is odd, the $Z_2$ fluxes obey sermionic statistics, while when $\mathcal K$ is even, the statistics of the $Z_2$ fluxes can be either fermionic or bosonic depending on the value of $n\in \mathbb Z$.

On the other hand, we recall that in two dimensions the $Z_2$ symmetry group protects two different SPT phases\cite{ChenGuLiuWen2011, ChenLiuWen, LevinGu, Else}, one is trivial and the other is nontrivial. Remarkably, in the nontrivial $Z_2$ SPT phase the $Z_2$ fluxes (by gauging the $Z_2$ symmetry) obey sermionic statistics while in the trivial SPT phase the $Z_2$ fluxes obey trivial bosonic statistics or fermionic statistics\cite{LevinGu}.

From the above discussion, we can conclude that the $\mathcal K=-1$ $SU(2)$ SPT phase corresponds to the $Z_2$ SPT phase when SOC is turned on, given that SOC reduces the $SU(2)$ symmetry to $Z_2$ (or a discrete group containing $Z_2$ as a subgroup) and that the resultant topological term flows to $\theta=2\pi$ (namely in the class $\mathcal K=-1$). Although spin Hall conductance is no longer well defined owing to the absence of conserved spin, the $Z_2$ symmetry can still protect anomalous edge excitations\cite{ChenLiuWen, ChenWen, LevinGu}.

	\section{Conclusions and discussions}  \label{sec:summary}
	
	In summary, we proposed a microscopic lattice model to realize the Hopf term in 2D spin systems. In our scenario, the spins couples to gapless Dirac fermions with both spin and orbit degrees of freedom, where the orbit degrees of freedom play an important role. The key point is that the magnetic moment of the spin is sensitive to the angular momentum $l_z$ of the orbitals of the electron: if $l_z=1$, namely if the orbital is in the state $p_x+ip_y$, then the electron feels a magnetic momentum parallel to $\pmb n$, but if $l_z=-1$, then the electron will feel a magnetic momentum anti-parallel to $\pmb n$ where $\pmb n$ is a smooth function of space and time. We also discussed the case in which the Dirac fermions have a small gap before coupling to the spins. In this case the value of $\theta$ is not quantized but we argue that under RG it will flow to a nearby fixed point which is quantized. The resultant ground state is an SPT phase protected by $SU(2)$ or $SO(3)$ symmetry. We further show that when spin-orbit coupling is considered an SPT phase protected by a discrete symmetry group can be obtained. This scenario might be generalized to realize other topological terms associated with the nontrivial mappings from space-time to spheres, for instance, $\pi_4(S^2)=\pi_4(S^3)=\mathbb Z_2$. 
	
	Our model shed light on the experimental realization of the Hopf topological term and the corresponding SPT phases. The most challenging part of experimental implimentation is that the coupling between the fermions and the spins depends on the status of the orbitals of the fermion.

	{\it Acknowledgement} We thank Yong Wang, Zhen-Yuan Yang, Meng Zhang, Qiang Luo, Xin Liu, Xiong-Jun Liu, Yu-Bin Li and Zheng-Cheng Gu for helpful discussions.  This work is supported by the NSF of China (Grants No. 11974421 and No. 12134020) and the National Key Research and Development Program of China (Grant No. 2022YFA1405301).

	\bibliography{Hopf_Latt_6}
	
	\widetext
	\appendix
	
	\section{Derivation of the Hopf term from a single Dirac cone}\label{app:singlecone}
	
	We start from the following relativistic action,
	\Beq
	 S_{\rm 0}=\int dx^{3}\bar{\psi}(i\gamma^{\mu}\partial_{\mu}+m\hat{n})\psi,
	\Eeq
	where $\hat n=\pmb n\cdot\pmb\sigma$ and $\gamma^\mu$ satisfy $\{\gamma^\mu,\gamma^\nu\} = g^{\mu\nu}$. The quantity in the integrant is the Lagrangian density $\mathcal L=\bar{\psi}(i\gamma^{\mu}\partial_{\mu}+m\hat{n})\psi$.
	
	From the Grassman formula
	\Beq
	e^{i S_{\rm eff}}=\int\mathcal{D}\bar{\psi}\mathcal{D}\psi e^{i\int d^{3}x\mathcal{L}}
	=\det(i\gamma^{\mu}\partial_{\mu}+m\hat n)
	\Eeq
	we have
	\Beq
	 S_{\rm eff}&=&-i\ln \det(i\gamma^{\mu}\partial_{\mu}+m\hat n)\\
	&=&-i{\rm Tr}\ln (i\gamma^{\mu}\partial_{\mu}+m\hat n)\label{e}\\ 
	&=&-i{\rm Tr}\ln D,
	\Eeq
	where we have define $D=i\gamma^{\mu}\partial_{\mu}+m\hat n$. Further defining $\bar D=-i\gamma^{\mu}\partial_{\mu}+m\hat n$, it follows that $$D\bar D=\partial^{2}+m^{2}+im\gamma^{\mu}(\partial_{\mu}\hat n),$$ where the bracket means that the differential operator $\partial_\mu$ only acts on $\hat n$. In latter discussion we will remove this bracket without causing confusion. The inverse of $D\bar D$ can be expanded in polynomial series of ${1\over\partial^2 + m^2}$ as the following,
	\Beq
	\frac{1}{D\bar D}&=&\frac{1}{\partial^{2}+m^{2}+im\gamma^{\mu}\partial_{\mu}\hat n}\\
	&=&\frac{1}{\partial^{2}+m^{2}}\Big[1+\sum_{k=1}^{\infty}\Big(-im\gamma^{\mu}\partial_{\mu}\hat n\frac{1}{\partial^{2}+m^{2}}\Big)^{k}\Big].
	\Eeq

	Introducing the variance $\delta \hat n=\delta \pmb n\cdot \pmb \sigma$, then we have $\delta D=m\delta \hat{n}$. The variance of the effective action reads 
	\beq\label{delS}
	\delta S_{\rm eff}(\pmb{n})&=&-i{\rm  Tr}(\delta DD^{-1})\notag\\
	&=&-i{\rm Tr}[\delta D\bar D(D\bar D)^{-1}]\notag\\
	&=&-i{\rm Tr}\Big[m\delta\hat n(-i\gamma^{\mu}\partial_{\mu}+m\hat n)\frac{1}{\partial^{2}+m^{2}}\Big(1+\sum_{k=1}^{\infty}\big(-im\gamma^{\mu}\partial_{\mu}\hat n\frac{1}{\partial^{2}+m^{2}}\big)^{k}\Big)\Big].
	\eeq
	
	In the above calculations, we have regarded $m$ as a large quantity. Hence in  later discussion we will only keep the lowest order of $m$ both in the dynamic terms (to the power of $m^1$) and in the topological term( to the power of $m^0)$\cite{Abanov, AbanovWiegmann}.

	Since the further calculations will frequently estimate the trace of $({1\over\partial^2 + m^2})^n$ over continuous indices, we first prove a general result for $N\geq2$
	\beq\label{tracen}
	{\rm Tr} \Big(\frac{1}{\partial^{2}+m^{2}}\Big)^{N} = \frac{i |m^{3-2N}|} {4(2N-3)\pi} {(2n-3)!!\over (2N-2)!!}.
	\eeq
	The trace can be calculated in momentum space,
	\Beq
	{\rm Tr} \Big(\frac{1}{\partial^{2}+m^{2}}\Big)^{N}
	&=&i\int dx^{3} \big(\frac{1}{\partial^{2}+m^{2}}\big)^{N}\\
	&=&i\int\frac{d^{3}\pmb{p}}{(2\pi)^{3}}\frac{1}{(\pmb{p}^{2}+m^{2})^{N}}\\
	&=&i\int_{0}^{\infty}\frac{dp}{2\pi^{2}}\frac{p^{2}}{(p^{2}+m^{2})^{N}}\\
	&=&\frac{i|m^{3-2N}|}{2\pi^{2} }\int_{0}^{\frac{\pi}{2}}d\theta \sin^{2}\theta \cos^{2N-4}\theta \\
	&=&\frac{i|m^{3-2N}|}{2(2N-3)\pi^{2}}\int_{0}^{\frac{\pi}{2}}d\theta \cos^{2N-2}\theta\\
	&=&\frac{i |m^{3-2N}|} {4(2N-3)\pi} {(2N-3)!!\over (2N-2)!!},
	\Eeq
	where we have used the formula
	\Beq
	I_{N}=\int_{0}^{\frac{\pi}{2}}d\theta \cos^{2N-2}\theta =\frac{2N-3}{2N-2}\cdot\frac{2N-5}{2N-4}\cdots\frac{3}{4}\cdot\frac{1}{2}\cdot\frac{\pi}{2} = {\pi\over2} {(2N-3)!!\over (2N-2)!!}.
	\Eeq

	Now we are ready to analyze the expansion term by term in Eq.(\ref{delS}).
	
	(A)Zeroth order:
	\Beq
	\delta S_{\rm eff}^{(0)}&=&-i{\rm Tr}\Big[m\delta\hat n(-i\gamma^{\mu}\partial_{\mu}+m\hat n)\frac{1}{\partial^{2}+m^{2}}\Big]\\
	&=&-{\rm Tr}\Big(m\delta\hat n\gamma^{\mu}\partial_{\mu}\frac{1}{\partial^{2}+m^{2}}\Big)-i{\rm Tr}\Big(m^{2}\delta\hat n\hat n\frac{1}{\partial^{2}+m^{2}}\Big)\\ \label{f}
	&=&-i{\rm Tr}\Big(m^{2}\delta\hat n\hat n\frac{1}{\partial^{2}+m^{2}}\Big) \\
	&=&-2im^{2}\delta n_{a}n^{a}{\rm Tr}\Big(\frac{1}{\partial^{2}+m^{2}}\Big) \label{g}\\
	&=&0\label{h}
	\Eeq
	where we have used ${\rm Tr\ } \pmb{\sigma}=0, {\rm Tr} (\sigma^{a}\sigma^{b})=2\delta^{ab}$ and $n_{a}n^{a}=|\pmb{n}|^2=1$.
	\\\\
	
	(B) First order:
	\Beq
	\delta S_{\rm eff}^{(1)}&=&-i{\rm Tr}\Big[m\delta\hat n(-i\gamma^{\mu}\partial_{\mu}+m\hat n)\frac{1}{\partial^{2}+m^{2}}\Big(-im\gamma^{\mu}\partial_{\mu}\hat n\frac{1}{\partial^{2}+m^{2}}\Big)\Big]\\
	&=&i{\rm Tr}\Big[m^{2}\delta\hat n\gamma^{\mu}\partial_{\mu}\frac{1}{\partial^{2}+m^{2}}\gamma^{\mu}\partial_{\mu}\hat n\frac{1}{\partial^{2}+m^{2}}\Big] 
	-{\rm Tr}\Big[m^{3}\delta\hat n\hat n\frac{1}{\partial^{2}+m^{2}}(\gamma^{\mu}\partial_{\mu}\hat n\frac{1}{\partial^{2}+m^{2}})\Big]\\
	&=&i{\rm Tr}\Big[m^{2}\delta\hat n\gamma^{\mu}\partial_{\mu}\frac{1}{\partial^{2}+m^{2}}\gamma^{\nu}\partial_{\nu}\hat n\frac{1}{\partial^{2}+m^{2}}\Big]\label{i}\\
	&=&2im^{2}{\rm Tr}(\delta\hat n\partial_{\mu}\partial^{\mu}\hat n){\rm Tr}\Big(\frac{1}{\partial^{2}+m^{2}}\frac{1}{\partial^{2}+m^{2}}\Big)\label{j}\\
	&=&-2im^{2}{\rm Tr}(\partial_{\mu}\delta\hat{n}\partial^{\mu}\hat{n}){\rm Tr}\Big(\frac{1}{\partial^{2}+m^{2}}\frac{1}{\partial^{2}+m^{2}}\Big)\label{k}\\
	&=&\frac{|m|}{4\pi}\int dx^{3}{\rm Tr}(\partial_{\mu}\delta\hat{n}\partial^{\mu}\hat{n})\label{l}
	\Eeq
	where we have used ${\rm Tr}\gamma^{\mu}=0, {\rm Tr} (\gamma^{\mu}\gamma^{\nu})=2g^{\mu\nu}$.

	(C) Second order:
	\Beq
	\delta S_{\rm eff}^{(2)}&=&-i{\rm Tr}\Big[m\delta\hat n(-i\gamma^{\mu}\partial_{\mu}+m\hat n)\frac{1}{\partial^{2}+m^{2}}(-im\gamma^{\mu}\partial_{\mu}\hat n\frac{1}{\partial^{2}+m^{2}})^{2}\Big]\\
	&=&{\rm Tr}\Big[m^{3}\delta\hat n\gamma^{\mu}\partial_{\mu}\frac{1}{\partial^{2}+m^{2}}(\gamma^{\mu}\partial_{\mu}\hat n\frac{1}{\partial^{2}+m^{2}})^{2}\Big]\label{m}
	+i{\rm Tr}\Big[m^{4}\delta\hat n\hat n\frac{1}{\partial^{2}+m^{2}}(\gamma^{\mu}\partial_{\mu}\hat n\frac{1}{\partial^{2}+m^{2}})^{2}\Big]\\
	&=&i{\rm Tr}\Big[m^{4}\delta\hat n\hat n\frac{1}{\partial^{2}+m^{2}}(\gamma^{\mu}\partial_{\mu}\hat n\frac{1}{\partial^{2}+m^{2}})^{2}\Big]\\
	&=&-\frac{|m|}{16\pi}\int dx^{3}{\rm tr}\delta\hat{n}\hat{n}\partial_{\mu}\hat{n}\partial^{\mu}\hat{n}. \label{n}
	\Eeq

	(D) Third order:
	\Beq
	\delta S_{\rm eff}^{(3)}&=&-i{\rm Tr}\Big[m\delta\hat n(-i\gamma^{\mu}\partial_{\mu}+m\hat n)\frac{1}{\partial^{2}+m^{2}}(-im\gamma^{\mu}\partial_{\mu}\hat n\frac{1}{\partial^{2}+m^{2}})^{3}\Big]\\
	&=&-i{\rm Tr}\Big[m^{4}\delta\hat n\gamma^{\mu}\partial_{\mu}\frac{1}{\partial^{2}+m^{2}}(\gamma^{\mu}\partial_{\mu}\hat n\frac{1}{\partial^{2}+m^{2}})^{3}\Big]\label{o} 
	+{\rm Tr}\Big[m^{5}\delta\hat n\hat n\frac{1}{\partial^{2}+m^{2}}(\gamma^{\mu}\partial_{\mu}\hat n\frac{1}{\partial^{2}+m^{2}})^{3}\Big]\\
	&=&{\rm Tr}\Big[m^{5}\delta\hat n\hat n\frac{1}{\partial^{2}+m^{2}}(\gamma^{\mu}\partial_{\mu}\hat n\frac{1}{\partial^{2}+m^{2}})^{3}\Big]\\
	&=&-\frac{{\rm sgn}(m)}{32\pi}\int dx^{3}\epsilon^{\mu\nu\rho} {\rm Tr}(\delta\hat{n}\hat{n}\partial_{\mu}\hat{n}\partial_{\nu}\hat{n}\partial_{\rho}\hat{n})\label{p}.
	\Eeq

	In summary
	\Beq
	\delta S_{\rm dyn}&=&\frac{|m|}{4\pi}\int dx^{3}{\rm Tr}(\partial_{\mu}\delta\hat{n}\partial^{\mu}\hat{n})+\cdots\\
	\delta S_{\rm topo}&=&-\frac{{\rm sgn}(m)}{32\pi}\int dx^{3}\epsilon^{\mu\nu\rho}{\rm Tr}(\delta\hat{n}\hat{n}\partial_{\mu}\hat{n}\partial_{\nu}\hat{n}\partial_{\rho}\hat{n})\label{im}
	\Eeq
	let $\hat{n}=2zz^{\dag}-1$,$z^{t}=(z_{1},z_{2})$-complex vector with unit modulus $z^{\dag}z = 1$.Substituting$\hat{n}=2zz^{\dag}-1$ into the above formula we obtain 
	\Beq
	\delta S_{\rm topo}&=&-\frac{{\rm sgn}(m)}{2\pi}\int dx^{3}\epsilon^{\mu\nu\rho}\delta a_{\mu}\partial_{\nu}a_{\lambda}
	\Eeq	
	where $a_{\mu}=z^{\dag}(-i\partial_{\mu})z$. Thus we obtain
	\Beq
	S_{\rm topo}&=&\pi\ {\rm sgn}(m) H(n)
	\Eeq	
	where $H(n)$ is the well-known Hopf invariant
	\Beq
	H(n)&=&-\frac{1}{4\pi^{2}}\int dx^{3}\epsilon^{\mu\nu\rho}a_{\mu}\partial_{\nu}a_{\lambda}
	\Eeq

	\section{Derivation of the low-energy effective action of the four-band model}\label{app:4band}
	
	In the four-band model, the coupling between the fermions and the decorated spins reads in momentum space,
	\Beq
	H_1 &=& \sum_{k,p} m  (C_{k A+}^\dag  \pmb \sigma  C_{pA+} -C_{k A-}^\dag  \pmb \sigma  C_{pA-} + C_{k B+}^\dag  \pmb \sigma  C_{pB+} - C_{k B-}^\dag  \pmb \sigma  C_{pB-}) \cdot \pmb n_{k-p}.
	\Eeq		
	
	Adopting the zero energy eigenstates at the cone $K'$, namely $\psi_{K'}^\dag(x) = [\psi^\dag_{K'A+}(x),\ \psi^\dag_{K'B-}(x)]$ as the bases of the low-energy subspace at the $K'$ valley, and expanding the Hamiltonian of the maintext and the above $H_1$ around the valley $K'$, one has
	\Beq
	\mathcal H_{K’+k} = v (k_x\gamma^x +k_y\gamma^y) + m \gamma_z\pmb n\cdot\pmb\sigma,
	\Eeq
	where $k$ is small and $v={\sqrt 3\over 4}(V_\pi-V_\sigma)=-{\sqrt 3\over 4}|(V_\pi-V_\sigma)|$.  So we have the effective Lagrangian 
	\Beq
	\mathcal L &=&  \psi_{K'}^{\dag}\Big[\big(i\partial_{t} -v(- i \gamma^{x}\partial_{x} - i\gamma^{y}\partial_{y})\big) - m\gamma^{z}\pmb n\cdot\pmb\sigma\Big]\psi_{K'} \notag\\
	&=& \psi_{K'}^{\dag}\Big[\big(i\partial_{t} -|v|( i \gamma^{x}\partial_{x} + i\gamma^{y}\partial_{y})\big) - m\gamma^{z}\pmb n\cdot\pmb\sigma\Big]\psi_{K'}.
	\Eeq

	Now we redefine $\bar\psi = \psi^\dag \gamma^0, \gamma^0=-\gamma^z, \gamma^1 = -\gamma^0\gamma^x, \gamma^2=-\gamma^0\gamma^y$, such that the $\gamma^\mu (\mu=0,1,2)$ matrices satisfy the relation 
	\Beq
	{\rm  Tr}(\gamma^{\mu}\gamma^{\nu})=2g^{\mu\nu}, 
	{\rm Tr}(\gamma^{\mu}\gamma^{\nu}\gamma^{\rho})=2i\epsilon^{\mu\nu\rho}
	\Eeq
	where $g={\rm diag}(1, -1, -1)$ is the three-dimensional Minkowski metric.

	By tuning the scale of space-time one can further set $|v|=1$, then the Lagrangian can be transformed into
	\beq\label{K2}
	\mathcal L_{K'} &=&  \bar\psi \Big[\big(i\gamma^0\partial_{t} + i\gamma^{1}\partial_{x} + i\gamma^{2}\partial_{y} \big) + m\pmb n\cdot\pmb\sigma\Big]\psi
	\eeq

	Similarly, around the $K$ valley, we adopt the bases $\psi_{K}^\dag(x) = [\psi^\dag_{KA-}(x),\ \psi^\dag_{KB+}(x)]$, 
	we have $\mathcal H_{K+k} = v (-k_x\gamma^x +k_y\gamma^y) + m\hat n$. After rescaling space-time we have
	\Beq
	\mathcal L_{K} &=&  \bar\psi_K \Big[\big(i\gamma^0\partial_{t} - i\gamma^{1}\partial_{x} + i\gamma^{2}\partial_{y} \big) - m\pmb n\cdot\pmb\sigma\Big]\psi_K
	\Eeq
	Now we change the bases as $\tilde \psi_{K}^\dag(x) = [\psi^\dag_{KB+}(x),\ -\psi^\dag_{KA-}(x)]$, then the Lagrangian takes the form
	\beq\label{K1}
	\mathcal L_{K} &=&  \bar{\tilde \psi}_K \Big[\big(i\gamma^0\partial_{t} +i\gamma^{1}\partial_{x} + i\gamma^{2}\partial_{y} \big) + m\pmb n\cdot\pmb\sigma\Big]\tilde \psi_K.
	\eeq

	Now the Lagrangian densities at $K$ (\ref{K1}) and $K'$ (\ref{K2}) are completely the same. Namely, we obtained two copies of Dirac fermions coupling to the same spin field with the same $m$.
	
	If we combine the bases together and introduce 
	\Beq
	\Psi_{k}^\dag&=&\Big( \psi^\dag_{K+k,B+\up}, \psi^\dag_{K+k,B+\dn}, -\psi^\dag_{K+k,A-\up}, -\psi^\dag_{K+k,A-\dn},\\
	&&\psi^\dag_{K'+k,A+\up}, \psi^\dag_{K'+k,A+\dn}, \psi^\dag_{K'+k,B-\up}, \psi^\dag_{K'+k,B-\dn} \Big).
	\Eeq 
	then we can write the action as
	\Beq
	S_{\rm eff}=\int dx^{3}\bar{\Psi}(i\gamma^{\mu}\partial_{\mu}+m\hat{n})\Psi,
	\Eeq
	where $\gamma^{0,1,2}$ are defined previously and we have rewriting $\hat{n}=\pmb {n}\cdot\pmb{\sigma}$.

	\section{No Hopf term from the coupling between two Dirac cones}\label{app:intercone}
	
	When ignoring the intercone coupling, the Lagrangian density reads,
	$ \mathcal L_0 = \bar \psi (D_0)   \psi, $
	where 
	\Beq
	D_{0}= i \gamma^\mu\partial_\mu + m\mathcal V_z\pmb n\cdot \pmb \sigma 
	\Eeq
	Defining $\bar D_0 =- i \gamma^\mu\partial_\mu + m\mathcal V_z\pmb n\cdot \pmb \sigma  $, then 
	\[
	\bar D_0D_0 = \partial^2  + m^2 - im\gamma^\mu\partial_\mu \hat n.
	\]
	The intercone coupling terms reads, 
	\Beq
	\mathcal{L}_{\rm int}&=&\pmb {n}_{K}(x) \cdot \bar{\psi}_{K^{'}}(x)\pmb{\sigma}\psi_{K}+\pmb{n}_{K}^{*}(x)\bar{\psi}_{K}\pmb{\sigma}\psi_{K^{'}}(x)\\
	&=&\bar{\psi}(\pmb{\lambda_{1}}\mathcal V_{x}+\pmb{\lambda_{2}}\mathcal V_{y})\cdot \pmb{\sigma}\psi
	\Eeq
	where $\pmb n_{K'}=\pmb n_K^*$, and $\pmb \lambda_{1}$ and $\pmb \lambda_{2}$ are the real and imaginary part of $\pmb n_K$. 
	We further define $D=D_0+(\pmb{\lambda_{1}}\mathcal V_{x}+\pmb{\lambda_{2}}\mathcal V_{y})\cdot \pmb{\sigma}$, and treat the intercone coupling as perturbation and only keep the linear term with respect to $\pmb n_K$, then we have
	\Beq
	\ln D&=&\ln(D_{0}+\pmb{\lambda_{1}}\mathcal V_{x}\pmb{\sigma}+\pmb{\lambda_{2}}\mathcal V_{y}\pmb{\sigma})\\
	&=&\ln D_{0}(1+\pmb{\lambda_{1}}D_{0}^{-1}\mathcal V_{x}\pmb{\sigma}+\pmb{\lambda_{2}}D_{0}^{-1}\mathcal V_{y}\pmb{\sigma})\\
	&=&\ln D_{0}+\ln(1+\pmb{\lambda_{1}}D_{0}^{-1}\mathcal V_{x}\pmb{\sigma}+\pmb{\lambda_{2}}D_{0}^{-1}\mathcal V_{y}\pmb{\sigma})\\
	&=&\ln D_{0}+\pmb{\lambda_{1}}D_{0}^{-1}\mathcal V_{x}\pmb{\sigma}+\pmb{\lambda_{2}}D_{0}^{-1}\mathcal V_{y}\pmb{\sigma}+O(\lambda^{2})
	\Eeq
	and,
	\Beq
	i S_{\rm eff} = {\rm Tr}(\ln D)\sim {\rm Tr}(\ln D_{0}) + {\rm Tr}(\pmb{\lambda_{1}}D_{0}^{-1}\mathcal V_{x}\pmb{\sigma}+\pmb{\lambda_{2}}D_{0}^{-1}\mathcal V_{y}\pmb{\sigma})+O(\lambda^{2}).
	\Eeq
	Since 
	\Beq
	{\rm Tr} (\pmb{\lambda_{1}}D_{0}^{-1}\mathcal V_{x}\pmb{\sigma}+\pmb{\lambda_{2}}D_{0}^{-1}\mathcal V_{y}\pmb{\sigma})
	= {\rm Tr}(\pmb{\lambda_{1}}(\bar D_{0}D_{0})^{-1}D_{0}^{+}\mathcal V_{x}\cdot \pmb{\sigma}+\pmb{\lambda_{2}}(\bar D_{0}D_{0})^{-1}\bar D_{0}\mathcal V_{y}\cdot \pmb{\sigma})\label{z},
	\Eeq
	both $(\bar D_{0} D_{0})^{-1}$ and $\bar D_{0}$ only contain $\mathcal V_0$ and $\mathcal V_z$ terms, therefore the trace in the linear expansion vanishes. 
	
	The higher order expansions do not vanish, but the terms which may contribute to the topological term of $\pmb n$ have negative power of $m$ (not shown), so are of no importance.

		\section{Hopf term from one massive Dirac cone}\label{app:m0}
	In this Appendix we provide details calculations for decorated spins coupling to fermions with a single massive Dirac cone. We consider the following action,
	\Beq
	S_{\rm 0}=\int dx^{3}\bar{\psi}(i\gamma^{\mu}\partial_{\mu}+m\hat{n} + m_0) \psi,
	\Eeq
	
	Defining $D=i\gamma^{\mu}\partial_{\mu}+m\hat n+m_{0}, \bar D=-i\gamma^{\mu}\partial_{\mu}+m\hat n+m_{0}$ and $M^2=m^2+m_0^2$, then we have 
	\Beq
	D\bar D=\partial^{2}+M^{2}+2m_{0}m\hat n+im\gamma^{\mu}\partial_{\mu}\hat n
	\Eeq
	and
	\Beq
	\frac{1}{D\bar D}&=&\frac{1}{\partial^{2}+M^{2}+2m_{0}m\hat n+im\gamma^{\mu}\partial_{\mu}\hat n}\\
	&=&\frac{1}{\partial^{2}+M^{2}}\Big[1+\sum_{k=1}^{\infty}(-2m_{0}m\hat n-im\gamma^{\mu}\partial_{\mu}\hat n)^{k}\big(\frac{1}{\partial^{2}+M^{2}}\big)^{k}\Big].
	\Eeq

	The effective action is given by $ S_{\rm eff}=-i{\rm Tr}\ln D$, and the variance of $S_{\rm eff}$ reads
	\Beq
	\delta S_{\rm eff}(\pmb{n})&=&-i{\rm Tr}(\delta DD^{-1})\\
	&=&-i{\rm Tr}[\delta D\bar D(D\bar D)^{-1}]\\
	&=&-i{\rm Tr}\Big[m\delta\hat n(-i\gamma^{\mu}\partial_{\mu}+m\hat n+m_{0})\frac{1}{\partial^{2}+M^2+2m_{0}m\hat n+im\gamma^{\mu}\partial_{\mu}\hat n}\Big]\\	
	&=&-i{\rm Tr}\Big\{m\delta\hat n(-i\gamma^{\mu}\partial_{\mu}+m\hat n+m_{0})\frac{1}{\partial^{2}+M^2}\Big[1+\sum_{k=1}^{\infty}(-2m_{0}m\hat n-im\gamma^{\mu}\partial_{\mu}\hat n)^{k}\big(\frac{1}{\partial^{2}+M^2}\big)^{k}\Big]\Big\}.
	\Eeq
	
	Now we  ignore the dynamical terms and only consider the terms which may generate the topological term. The lowest order which contribute to the Hopf term appear at $k=3$,
	\Beq
	\delta S_{\rm eff}^{(3)}&=&-i{\rm Tr}\Big[m\delta\hat n(-i\gamma^{\mu}\partial_{\mu}+m\hat n+m_{0})\frac{1}{\partial^{2}+M^2}(-2m_{0}m\hat n-im\gamma^{\mu}\partial_{\mu}\hat n)^{3}\Big(\frac{1}{\partial^{2}+M^2}\Big)^{3}\Big]\\
	&=&-\frac{m^{5}}{32\pi(\sqrt{M^2})^{5}}\int dx^{3}\epsilon^{\mu\nu\rho}{\rm tr}(\delta\hat{n}\hat{n}\partial_{\mu}\hat{n}\partial_{\nu}\hat{n}\partial_{\rho}\hat{n})\label{p}.
	\Eeq
	When $m_0=0$ the above formula reproduces the previous result. 
	
	When $m_0\neq 0$, there will be many terms in the expansion which may have contribution to the Hopf term. We only  consider the terms in which the power of $m$ and $m_0$ are both zero.  For $k\geq3$, we have
	
	\Beq
		\delta S_{\rm eff}^{(k)}=&-i{\rm Tr}[m\delta\hat n(-i\gamma^{\mu}\partial_{\mu}+m\hat n+m_{0})\frac{1}{\partial^{2}+M^2}(-2m_{0}m\hat n-im\gamma^{\mu}\partial_{\mu}\hat n)^{k	}(\frac{1}{\partial^{2}+M^2})^{k}].\label{D1}
	\Eeq
	
	Since $\partial_\mu\hat{n}^{2}=0$, $\hat{n}$ anticommutes with $\partial_\mu\hat{n}$. 
Therefore the calculation of $(-2m_{0}m\hat n-im\gamma^{\mu}\partial_{\mu}\hat n)^{k}$ in the term cannot be treated as a general binomial expansion. To get the topological terms, we have to pick three $(-im\gamma^{\mu}\partial_{\mu}\hat n)$ terms in the expansion of the polynomial. 
Due to the anticommuting relation, many intermediate terms cancel with each other.
When $k$ is even, there are no such terms; and when $k$ is odd, the number of such terms is $\frac{k-1}{2}$.

Defining $x={m_0\over m}$, then we have,
	\Beq
	\delta S_{\rm topo}^{(k)} &=& -\frac{m_{0}^{k-3}m^{k+2}}{(\sqrt{M^2})^{2k-1}}(\frac{2^{k-4}(k-1)}{4k\pi}\cdot\frac{2k-3}{2k-2}\cdot\frac{2k-5}{2k-4}\cdots\frac{3}{4}\cdot\frac{1}{2})\int dx^{3}\epsilon^{\mu\nu\rho}{\rm tr}(\delta\hat{n}\hat{n}\partial_{\mu}\hat{n}\partial_{\nu}\hat{n}\partial_{\rho}\hat{n})\\
	&=& - {\rm sgn}(m)(\frac{2x}{1+x^2})^{k-3}\cdot\frac{1}{(\sqrt{1+x^2})^5}(\frac{(k-1)}{8k\pi}\cdot\frac{2k-3}{2k-2}\cdot\frac{2k-5}{2k-4}\cdots\frac{3}{4}\cdot\frac{1}{2})\int dx^{3}\epsilon^{\mu\nu\rho}{\rm tr}(\delta\hat{n}\hat{n}\partial_{\mu}\hat{n}\partial_{\nu}\hat{n}\partial_{\rho}\hat{n}).
	\Eeq
	
Introducing a family of functions of $x$
	\Beq
	f_k(x) &=& - {\rm sgn}(m){x^{k-3}\over   \big(\sqrt{1+x^2}\big)^{2k-1}}(\frac{2^{k-4}(k-1)}{4k\pi}\cdot\frac{2k-3}{2k-2}\cdot\frac{2k-5}{2k-4}\cdots\frac{3}{4}\cdot\frac{1}{2}),
	\Eeq
	then 
\Beq
	\delta S_{\rm topo}=\sum_{N=1}^{\infty}  f_{2N+1} \int dx^{3}\epsilon^{\mu\nu\rho} {\rm tr}(\delta\hat{n}\hat{n}\partial_{\mu}\hat{n}\partial_{\nu}\hat{n}\partial_{\rho}\hat{n}),
\Eeq
	and consequently
\beq
	 S_{\rm topo} &=& -{\theta \over 4\pi^2}\int d^x \epsilon^{\mu\nu\rho}a_\mu\partial_\nu a_\rho,\\
	\theta(x) &=& - 32\pi^2 \sum_{N=1}^{\infty} f_{2N+1}.\label{thetax}
\eeq
	The convergence of the series requires that
	\Beq
	\lim_{N\rightarrow\infty}\frac{f_{2N+1}}{f_{2N-1}}=\frac{4m_{0}^{2}m^{2}}{(\sqrt{M^2})^{2}}={4x^2\over (1+x^2)^2}<1.
	\Eeq
When $x^2=({m_0\over m})^2=1$, the series diverges. Away from $x=\pm1$, the series converges. Especially, 
	\Beq
	&&\theta(0) = {\pi}\ {\rm sgn }(m), \\
	&&\theta(\infty) = {0}.
	\Eeq
	
	\section{Hopf term from SOC}\label{app:SOC}
	
	Considering the Hamiltonian from the case of spin-orbit coupling, the original spin-independent Hamiltonian matrix becomes spin-dependent. The original 4 × 4 matrix is expanded to an 8 × 8 matrix, which is also an 8 × 8 matrix,  with spin simplification. Consider the Hamiltonian matrix for the case of adding the SOC as\cite{WuCJ2}
	\Beq
		H^{\sigma\sigma}_{\rm eff}=
		\left(\begin{matrix} 
			H^{\sigma\sigma}_{\uparrow\uparrow} &	0 \\0 &H^{\sigma\sigma}_{\downarrow\downarrow} 
		\end{matrix}\right)
	\Eeq
	where the 4 × 4 matrix on the diagonal is as follows
	$$H^{\sigma\sigma}_{\uparrow\uparrow} =\left(\begin{array}{cccc}
		0&-i\lambda_{\rm so}&h_{xx}^{AB}&h_{xy}^{AB}\\
		i\lambda_{\rm so}&0&h_{yx}^{AB}&h_{yy}^{AB}\\
		h_{xx}^{\dag AB}&h_{yx}^{\dag AB}&0&-i\lambda_{\rm so}\\
		h_{xy}^{\dag AB}&h_{yy}^{\dag AB}&i\lambda_{\rm so}&0
	\end{array}\right)$$
	
	$$H^{\sigma\sigma}_{\downarrow\downarrow}= \left(\begin{array}{cccc}
		0&i\lambda_{\rm so}&h_{xx}^{AB}&h_{xy}^{AB}\\
		-i\lambda_{\rm so}&0&h_{yx}^{AB}&h_{yy}^{AB}\\
		h_{xx}^{\dag AB}&h_{yx}^{\dag AB}&0&i\lambda_{\rm so}\\
		h_{xy}^{\dag AB}&h_{yy}^{\dag AB}&-i\lambda_{\rm so}&0
	\end{array}\right)$$
	
The matrix elements here are as follows
	\Beq
	&&h_{xx}^{AB}=V_{n}+\frac{1}{2}(3V_{\pi}+V_{\sigma})e^{i\frac{\sqrt{3}}{2}k_{y}}\cos\frac{k_{x}}{2},\\
	&&h_{xy}^{AB}=h_{yx}^{AB}=i\frac{\sqrt{3}}{2}(V_{\pi}-V_{\sigma})e^{i\frac{\sqrt{3}}{2}k_{y}}\sin\frac{k_{x}}{2},\\
	&&h_{yy}^{AB}=V_{\pi}+\frac{1}{2}(V_{\pi}+3V_{\sigma})e^{i\frac{\sqrt{3}}{2}k_{y}}\cos\frac{k_{x}}{2}.
	\Eeq
	
	The $\lambda_{\rm so}$ term arises from the intrinsic SOC $\pmb L\cdot\pmb S$. 
Since the value of $\lambda_{\rm so}$ is small (the values of $\lambda_{\rm so}$ for some specific materials are shown in the table \ref{clcs}), we can consider it as perturbation. This allows us to project it to the Hamiltonian matrix $H^{\sigma\sigma}_{0}$ at no mass
	on the four eigenstates with zero eigenvalues. After calculation, the matrix form of the projection operator at $K^{'}$ can be written as
	\Beq
		P=
		\left(\begin{matrix} 
			-i&0&0&0\\
			1&0&0&0\\
			0&i&0&0\\
			0&1&0&0\\
			0&0&-i&0\\
			0&0&1&0\\
			0&0&0&i\\
			0&0&0&1	
		\end{matrix}\right).
	\Eeq
	Projecting onto the middle four energy bands, yielding
		\Beq
		\left(\begin{matrix} 
			2&0&0&0\\
			0&-2&0&0\\
			0&0&-2&0\\
			0&0&0&2
		\end{matrix}\right).\lambda_{\rm so}
		\Eeq
	
Note that the base of the matrix at this moment is $|1 \uparrow\rangle, |2 \uparrow\rangle, |1 \downarrow\rangle, |2 \downarrow\rangle$, where 1,2 are the energy band indices (which involve the Dirac cones) mixing the sublattice indices (A,B) and the orbit idices ($p_{x},p_{y}$).$\uparrow$,$\downarrow$ are the spin indicators. We can write the above matrix in a more explicit form
	$$\sigma^{z}\gamma^{z}\lambda_{\rm so}$$
	where $\sigma^{z},\gamma^{z}$ are both Pauli matrices acting on the spin and energy bands, respectively. On the other hand, it is calculated that the projection operator at $K$ has the following matrix form:
	\Beq
		P=
		\left(\begin{matrix} 
			i&0&0&0\\
			1&0&0&0\\
			0&-i&0&0\\
			0&1&0&0\\
			0&0&i&0\\
			0&0&1&0\\
			0&0&0&-i\\
			0&0&0&1	
		\end{matrix}\right).
	\Eeq
	Similarly, projecting to the middle four energy bands yields:
	\Beq
		\left(\begin{matrix} 
			-2&0&0&0\\
			0&2&0&0\\
			0&0&2&0\\
			0&0&0&-2
		\end{matrix}\right).\lambda_{\rm so}
	\Eeq
	We write the above matrix as:
	$$
	-\sigma^{z}\gamma^{z}\lambda_{\rm so}.
	$$

Since the coupling between the two cones can be ignored, the topological term contributed by the cones can be calculated separately. We consider the valley $K'$ with Lagrangian
	\begin{align}
		\mathcal L_{K'} =&  \bar\psi \Big[\big(i\gamma^0\partial_{t} + i\gamma^{1}\partial_{x} + i\gamma^{2}\partial_{y} \big) + m\pmb n\cdot\pmb\sigma\notag
		+\sigma^{z}\lambda_{\rm so}.
	\end{align}
	
Defining $\gamma$ and  $D$
		\Beq
		\bar\psi &=& \psi^\dag \gamma^0, \gamma^0=-\gamma^z, \gamma^1 = -\gamma^0\gamma^x, \gamma^2=-\gamma^0\gamma^y\\
		D&=&i\gamma^{\mu}\partial_{\mu}+m\hat{n}	+\sigma^{z}\lambda_{\rm so},
		\bar{D}=-i\gamma^{\mu}\partial_{\mu}+m\hat{n}+\sigma^{z}\lambda_{\rm so}
		\Eeq

and $M^2=m^2+\lambda_{\rm so}^2$, one has

	\Beq
		D\bar D=&\partial^{2}+M^{2}+\lambda_{\rm so}m\{\sigma^{z},\hat n\}+im\gamma^{\mu}\partial_{\mu}\hat n.
	\Eeq
	
	Because the path integration process requires the trace of the operator, the inverse of the operator $\frac{1}{D\bar D}$ is used, which can be expanded by order as
	\Beq
		\frac{1}{D\bar D}&=&\frac{1}{\partial^{2}+M^{2}+\lambda_{\rm so}m{\sigma^{z},\hat n}+im\gamma^{\mu}\partial_{\mu}\hat n}\\
		&=&\frac{1}{\partial^{2}+M^{2}}\Big[1+\sum_{k=1}^{\infty}(-\lambda_{\rm so}m\{\sigma^{z},\hat n\}-im\gamma^{\mu}\partial_{\mu}\hat n)^{k}\big(\frac{1}{\partial^{2}+M^{2}}\big)^{k}\Big].
	\Eeq
	
	According to the previous calculation, the effective action amount is
	$S_{\rm eff}=-i{\rm Tr}\ln D$
	.Doing the variation on the field $\hat{n}$, $\delta\hat{n}=\delta\pmb{n}\cdot\pmb{\sigma}$, the variation of $S_{\rm eff}$ is written as
	\Beq
		\delta S_{\rm eff}(\pmb{n})&=&-i{\rm Tr}(\delta DD^{-1})\\
		&=&-i{\rm Tr}[\delta D\bar D(D\bar D)^{-1}]\\
		&=&-i{\rm Tr}\Big[m\delta\hat n(-i\gamma^{\mu}\partial_{\mu}+m\hat n+\sigma^{z}\lambda_{\rm so})\frac{1}{\partial^{2}+M^2+\lambda_{\rm so}m\{\sigma^{z},\hat n\}+im\gamma^{\mu}\partial_{\mu}\hat n}\Big]\\	
		&=&-i{\rm Tr}\Big\{m\delta\hat n(-i\gamma^{\mu}\partial_{\mu}+m\hat n+\sigma^{z}\lambda_{\rm so})\frac{1}{\partial^{2}+M^2}\Big[1+\sum_{k=1}^{\infty}(-\lambda_{\rm so}m\{\sigma^{z},\hat n\}-im\gamma^{\mu}\partial_{\mu}\hat n)^{k}\big(\frac{1}{\partial^{2}+M^2}\big)^{k}\Big]\Big\}.
	\Eeq
	
	Now we leave the dynamical terms aside for the moment and consider only the expansion terms that may generate topological terms. Similarly, the lowest order expansion term contributing to the Hopf term appears at $k=3$, which is consistent with the order of the Hopf term appearing in the previous Hopf model as well as the four-energy band model.
	\Beq
		\delta S_{\rm eff}^{(3)}&=&-i{\rm Tr}\Big[m\delta\hat n(-i\gamma^{\mu}\partial_{\mu}+m\hat n+\sigma^{z}\lambda_{\rm so})\frac{1}{\partial^{2}+M^2}(-\lambda_{\rm so}m\{\sigma^{z},\hat n\}-im\gamma^{\mu}\partial_{\mu}\hat n)^{3}\Big(\frac{1}{\partial^{2}+M^2}\Big)^{3}\Big]\\
		&=&-\frac{m^{5}}{32\pi(\sqrt{M^2})^{5}}\int d^{3}x\epsilon^{\mu\nu\rho}{\rm tr}(\delta\hat{n}\hat{n}\partial_{\mu}\hat{n}\partial_{\nu}\hat{n}\partial_{\rho}\hat{n}).
	\Eeq
	
	It is worth noting that when $\lambda_{\rm so}=0$, the above equation is exactly the result when the Dirac fermions are massless. This is physically self-consistent, since it corresponds to the absence of spin-orbit coupling, which should then return to the massless case.
	
	When $\lambda_{\rm so}\neq 0$, there are an infinite number of possible contributions to generate Hopf terms in the expansion. We only consider terms with zero total powers of $m$ and $\lambda_{\rm so}$. This is because these terms do not change and do not converge to zero as $m$ flows to large values. Next, computing the $k\geq3$ case, each $k$-order expansion term can be written as
	\begin{align}
		\delta S_{\rm eff}^{(k)}=-i{\rm Tr}[m\delta\hat n(-i\gamma^{\mu}\partial_{\mu}+m\hat n+\sigma^{z}\lambda_{\rm so})\frac{1}{\partial^{2}+M^2}(-\lambda_{\rm so}m\{\sigma^{z},\hat n\}-im\gamma^{\mu}\partial_{\mu}\hat n)^{k}(\frac{1}{\partial^{2}+M^2})^{k}]\label{gogo}
	\end{align}
	
	Unlike the previous case where a constant mass $m_{0}$ was added, this time the expansion terms of $(-\lambda_{\rm so}m\{\sigma^{z},\hat n\}-im\gamma^{\mu}\partial_{\mu}\hat n)^{k}$ have the non-commutability of $\sigma^{z}$ and $\hat n$ in addition to the anticommutation of $\hat{n}$ and $\partial_{\mu}\hat{n}$. We write $\sigma^{z} \hat{n} = \hat{n}^{*}\sigma^{z}$, where $\hat{n}^{*} =-n_{x}\sigma^{x}-n_{y}\sigma^{y}+n_{z}\sigma^{z}$. Continuing to examine Eq. (\ref{gogo}), to get the topological term we need to pick three ($-im\gamma^{\mu}\partial_{\mu}\hat n$) and the remaining $(k-3)$ product factors are ($-\lambda_{\rm so}m\sigma^{z}\hat n$) or ($-\lambda_{\rm so}m\hat n\sigma^{z}$). We can think of ($-\lambda_{\rm so}m\sigma^{z}\hat n$) and ($-\lambda_{\rm so}m\hat n\sigma^{z}$) as positive and negative signs on a line, and it is the adjacent positive and negative signs that can cancel. We can cancel this whole line only if the number of ($-\lambda_{\rm so}m\sigma^{z}\hat n$) and ($-\lambda_{\rm so}m\hat n\sigma^{z}$) are equal.
	
	There is another limitation we have to consider, the insertion of the three ($-im\gamma^{\mu}\partial_{\mu}\hat n$) divides the line into four segments(the case of zero signs also counts as a segment). Due to  the non-commutability of $\hat{n}^{*}$ and $\partial_{\mu}\hat{n}$ ,we have to make the number of positive and negative signs in each segment equal to get the topological term.

	\begin{figure}
		\centering
		\includegraphics[width=0.5\linewidth]{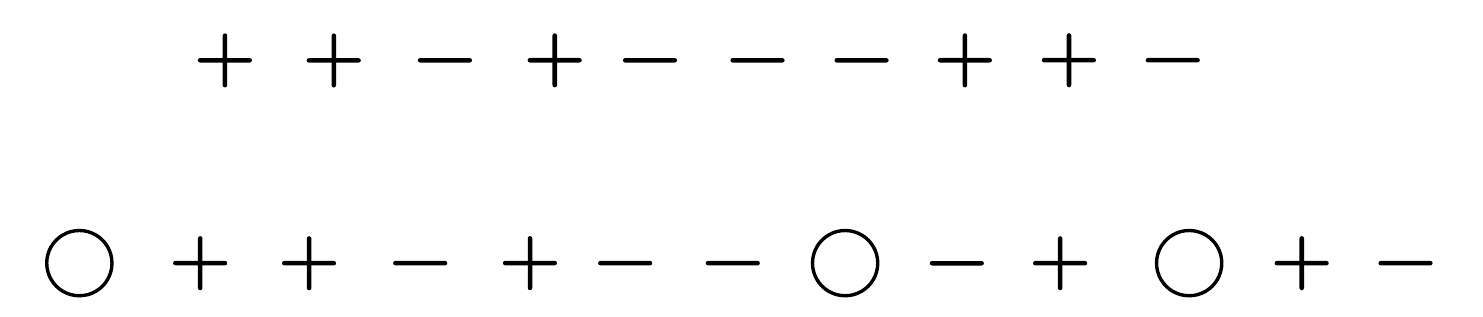}
		\caption[signs]{The positive and negative signs represent ($-\lambda_{\rm so}m\sigma^{z}\hat n$) and   ($-\lambda_{\rm so}m\hat n\sigma^{z}$), and the circles represent $(-im\gamma^{\mu}\partial_{\mu}\hat n)$.}
		\label{fig:signs}
	\end{figure}
	
	It can be shown that for each $k$, $m\hat n$ and $\sigma^{z}\lambda_{\rm so}$ in the first parenthesis cannot appear in the expansion at the same time. When $k$ is odd, $m\hat n$ appears in the final result, and when $k $ is even, $\sigma^{z}\lambda_{\rm so}$ appears. When $k>3$, the $\hat n$ matrix can be contracted in pairs since $\hat n^2=I$ (where $I$ is a 2$\times$2 unit matrix). We will discuss these cases separately, and here we define $x={\lambda_{\rm so}\over m}$.\\\\
	(1) When $k$ is odd, we have
	\Beq
		\delta S_{\rm topo}^{(k)} &=& -\frac{\lambda_{\rm so}^{k-3}m^{k+2}}{(\sqrt{M^2})^{2k-1}}\cdot\eta(k)\cdot(\frac{1}{4k\pi}\cdot\frac{2k-3}{2k-2}\cdot\frac{2k-5}{2k-4}\cdots\frac{3}{4}\cdot\frac{1}{2})\cdot\int d^{3}x\epsilon^{\mu\nu\rho}{\rm tr}(\delta\hat{n}\hat{n}\partial_{\mu}\hat{n}\partial_{\nu}\hat{n}\partial_{\rho}\hat{n})\\
		&=& - {\rm sgn}(m){x^{k-3}\over   \big(\sqrt{1+x^2}\big)^{2k-1}}\cdot\eta(k)\cdot(\frac{1}{4k\pi}\cdot\frac{2k-3}{2k-2}\cdot\frac{2k-5}{2k-4}\cdots\frac{3}{4}\cdot\frac{1}{2})\cdot\int d^{3}x\epsilon^{\mu\nu\rho}{\rm tr}(\delta\hat{n}\hat{n}\partial_{\mu}\hat{n}\partial_{\nu}\hat{n}\partial_{\rho}\hat{n})
	\Eeq
	
	where $\eta(k)$ is the number of expansion terms that meet the requirements.\\\\
	(2) When $k$ is even, we have
	\Beq
		\delta S_{\rm topo}^{(k)} &=&   \frac{\lambda_{\rm so}^{k-2}m^{k+1}}{(\sqrt{M^2})^{2k-1}}\cdot\eta(k)\cdot(\frac{1}{4k\pi}\cdot\frac{2k-3}{2k-2}\cdot\frac{2k-5}{2k-4}\cdots\frac{3}{4}\cdot\frac{1}{2})\cdot\int d^{3}x\epsilon^{\mu\nu\rho}{\rm tr}(\delta\hat{n}\hat{n}\partial_{\mu}\hat{n}\partial_{\nu}\hat{n}\partial_{\rho}\hat{n})\\
		&=& {\rm sgn}(m){x^{k-2}\over   \big(\sqrt{1+x^2}\big)^{2k-1}}\cdot\eta(k)\cdot(\frac{1}{4k\pi}\cdot\frac{2k-3}{2k-2}\cdot\frac{2k-5}{2k-4}\cdots\frac{3}{4}\cdot\frac{1}{2})\cdot\int d^{3}x\epsilon^{\mu\nu\rho}{\rm tr}(\delta\hat{n}\hat{n}\partial_{\mu}\hat{n}\partial_{\nu}\hat{n}\partial_{\rho}\hat{n})
	\Eeq
	Define the two families of functions
	\Beq
		f_k(x) &=&- {\rm sgn}(m){x^{k-3}\over   \big(\sqrt{1+x^2}\big)^{2k-1}}\cdot\eta(k)\cdot(\frac{1}{4k\pi}\cdot\frac{2k-3}{2k-2}\cdot\frac{2k-5}{2k-4}\cdots\frac{3}{4}\cdot\frac{1}{2})\\ 
		g_k(x)&=&+{\rm sgn}(m){x^{k-2}\over   \big(\sqrt{1+x^2}\big)^{2k-1}}\cdot\eta(k)\cdot(\frac{1}{4k\pi}\cdot\frac{2k-3}{2k-2}\cdot\frac{2k-5}{2k-4}\cdots\frac{3}{4}\cdot\frac{1}{2})
    \Eeq
	Then the summation of these infinite terms is the variation of the topological term in the effective action,
\Beq
		\delta S_{\rm topo}=\sum_{N=1}^{\infty}  [f_{2N+1}+g_{2N+2}] \int d^{3}x\epsilon^{\mu\nu\rho}tr(\delta\hat{n}\hat{n}\partial_{\mu}\hat{n}\partial_{\nu}\hat{n}\partial_{\rho}\hat{n}).
\Eeq
So the topological term can be calculated with
	\beq
		S_{\rm topo} &=& -{\theta \over 4\pi^2}\int d^{3}x \epsilon^{\mu\nu\rho}a_\mu\partial_\nu a_\rho,\\
		\theta(x) &=& - 32\pi^2 \sum_{N=1}^{\infty}  [f_{2N+1}+g_{2N+2}].\label{theta2}
	\eeq
	\begin{table} 
		\centering\large
		\begin{tabular}{|c|c|}		
			\hline
			Materials&$\lambda_{\rm so}(eV)$\\
			\hline	\hline
			Bi/SiC&0.435\\
			\hline
			Sb/SiC&0.2\\
			\hline
			As/SiC&0.006\\
			\hline		
		\end{tabular}
		\caption[Material parameters]{ Values of $\lambda_{\rm so}$ in various models\cite{WuCJ2}}\label{fig:lamr}\label{clcs}
	\end{table} 	
We can see that for the case of adding SOC, although the added terms are different, after the path integral, the expressions in the result of the topological term variance appear to be the same as the case of adding $m_{0}$ analyzed in the previous subsection, and then the topological terms computed are the same. Next, we analyze the effect of $\lambda_{\rm so}$ on the dynamical terms and compute (\ref{gogo}) term by term.\\	
	(A) Zeroth order term:
	\Beq
		\delta S_{\rm eff}^{(0)}&=&-i{\rm Tr}\Big[m\delta\hat n(-i\gamma^{\mu}\partial_{\mu}+m\hat n+\sigma^{z}\lambda_{\rm so})\frac{1}{\partial^{2}+M^{2}}\Big]\\
		&=&-{\rm Tr}\Big(m\delta\hat n\gamma^{\mu}\partial_{\mu}\frac{1}{\partial^{2}+M^{2}}\Big)-i{\rm Tr}\Big(m^{2}\delta\hat n\hat n\frac{1}{\partial^{2}+M^{2}}\Big)-i{\rm Tr}\Big(m\lambda_{\rm so}\sigma^{z}\delta\hat n\frac{1}{\partial^{2}+M^{2}}\Big)\\ 
		&=&-i{\rm Tr}\Big(m^{2}\delta\hat n\hat n\frac{1}{\partial^{2}+M^{2}}\Big) \\
		&=&-2im^{2}\delta n_{a}n^{a}{\rm Tr}\Big(\frac{1}{\partial^{2}+M^{2}}\Big) \\
		&=&0.
	\Eeq
The derivation here we use the relation 
	\Beq
		{\rm Tr\ } \pmb{\sigma}=0, {\rm Tr} (\sigma^{a}\sigma^{b})=2\delta^{ab}, n_{a}n^{a}=|\pmb{n}|^2=1.
	\Eeq
	(B) First-order term:
	\Beq
		\delta S_{\rm eff}^{(1)}&=&-i{\rm Tr}[m\delta\hat n(-i\gamma^{\mu}\partial_{\mu}+m\hat n+\sigma^{z}\lambda_{\rm so})\frac{1}{\partial^{2}+M^2}(-\lambda_{\rm so}m\{\sigma^{z},\hat n\}-im\gamma^{\mu}\partial_{\mu}\hat n)(\frac{1}{\partial^{2}+M^2})]\\
		&=&i{\rm Tr}\Big[m^{2}\delta\hat n\gamma^{\mu}\partial_{\mu}\frac{1}{\partial^{2}+M^{2}}\gamma^{\mu}\partial_{\mu}\hat n\frac{1}{\partial^{2}+M^{2}}\Big]-{\rm Tr}\Big[m^{3}\delta\hat n\hat n\frac{1}{\partial^{2}+M^{2}}(\gamma^{\mu}\partial_{\mu}\hat n\frac{1}{\partial^{2}+M^{2}})\Big]\notag\\
		&&-{\rm Tr}\Big[m^{2}\delta\hat n\lambda_{\rm so}\sigma^{z}\frac{1}{\partial^{2}+M^{2}}(\gamma^{\mu}\partial_{\mu}\hat n\frac{1}{\partial^{2}+M^{2}})\Big]+i{\rm Tr}\Big[\lambda_{\rm so}m^{3}\delta\hat n n_{z}\frac{1}{\partial^{2}+M^{2}}\frac{1}{\partial^{2}+M^{2}}\Big]\notag\\
		&&+i{\rm Tr}\Big[\lambda_{\rm so}^{2}m^{2}\delta\hat n \sigma_{z}\hat{n}\sigma_{z}\frac{1}{\partial^{2}+M^{2}}\frac{1}{\partial^{2}+M^{2}}\Big]\\
		&=&-2im^{2}{\rm Tr}(\partial_{\mu}\delta\hat{n}\partial^{\mu}\hat{n}){\rm Tr}\Big(\frac{1}{\partial^{2}+M^{2}}\frac{1}{\partial^{2}+M^{2}}\Big)+i{\rm Tr}\Big[\lambda_{\rm so}m^{3}\delta\hat n n_{z}\frac{1}{\partial^{2}+M^{2}}\frac{1}{\partial^{2}+M^{2}}\Big]\notag\\
		&&+i{\rm Tr}\Big[\lambda_{\rm so}^{2}m^{2}\delta\hat n \sigma_{z}\hat{n}\sigma_{z}\frac{1}{\partial^{2}+M^{2}}\frac{1}{\partial^{2}+M^{2}}\Big]\\
		&=&\frac{m^{2}}{4\pi M}\int d^{3}x{\rm tr}(\partial_{\mu}\delta\hat{n}\partial^{\mu}\hat{n})-\frac{\lambda_{\rm so}m^{3}}{8\pi M}\int d^{3}x{\rm tr}(\delta\hat{n}\hat{n}_{z})-\frac{\lambda_{\rm so}^{2}m^{2}}{8\pi M}\int d^{3}x{\rm tr}(\delta\hat{n}\sigma_{z}\hat{n}\sigma_{z}).
	\Eeq
	    \begin{figure}[t]
		\centering
		\includegraphics[width=0.5\linewidth]{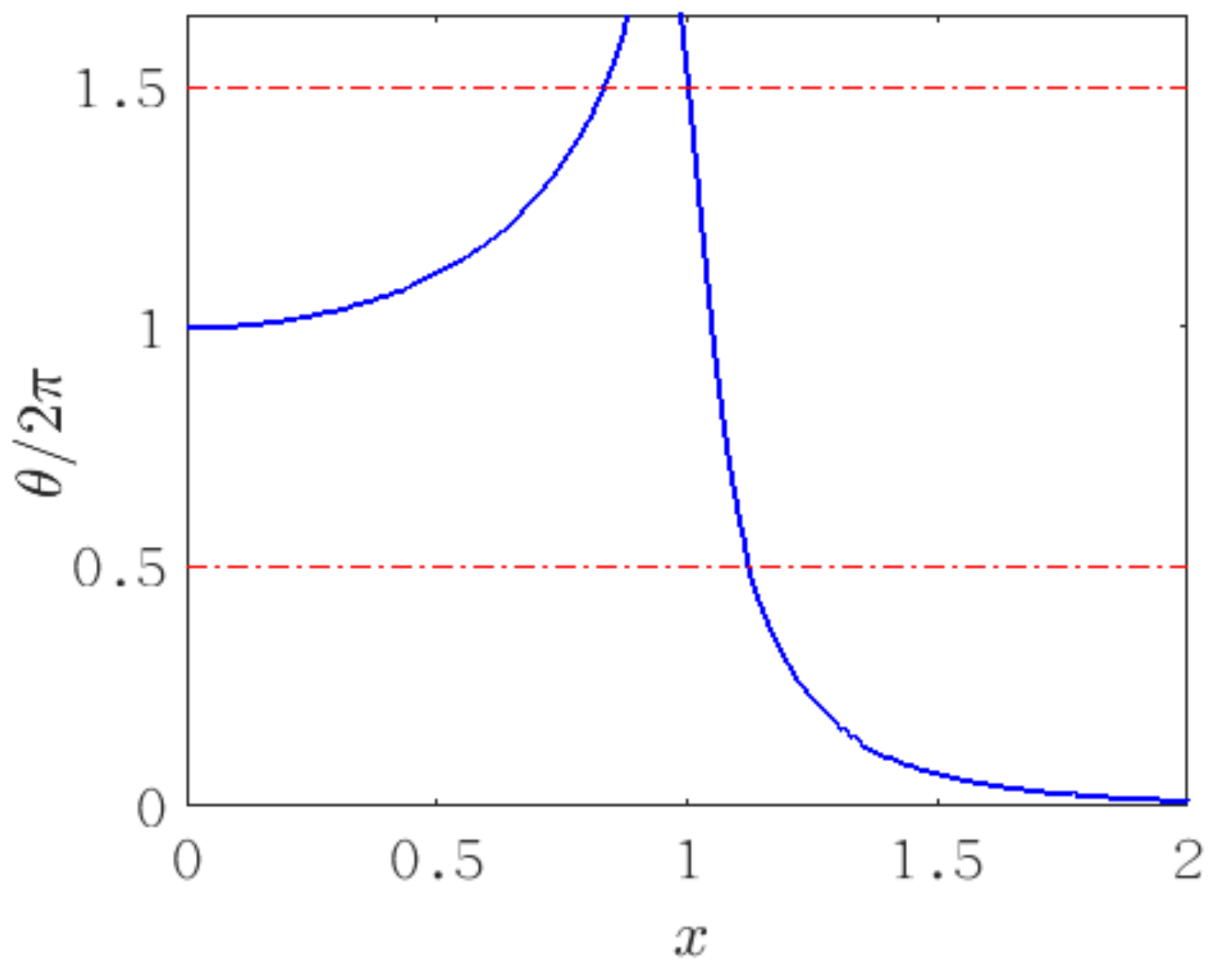}
		\caption[4band]{The value of $\theta(x)$ as a function of $x={{\lambda_{\rm so}}\over m}$ with $\theta(-x)=\theta(x)$. Here we have summed over the first $M=500$ terms in the series (\ref{theta2}). The function $\theta(x)$ diverges at $|x|=1$ with increasing $M$. Between the two dashed red lines, $\theta$ flows to $2\pi$ under RG.  
		} 
		\label{fig:SOC_500}
	\end{figure}
	The derivation here we use the relation 
	\Beq
		{\rm Tr}\gamma^{\mu}=0, {\rm Tr} (\gamma^{\mu}\gamma^{\nu})=2g^{\mu\nu}.
	\Eeq 
	(C) Second-order term:
	\Beq
		\delta S_{\rm eff}^{(2)}&=&-i{\rm Tr}[m\delta\hat n(-i\gamma^{\mu}\partial_{\mu}+m\hat n+\sigma^{z}\lambda_{\rm so})\frac{1}{\partial^{2}+M^2}(-\lambda_{\rm so}m\{\sigma^{z},\hat n\}-im\gamma^{\mu}\partial_{\mu}\hat n)^{2}(\frac{1}{\partial^{2}+M^2})^{2}]\\
		&=&i2m^{4}{\rm Tr}\Big[\delta\hat{n}\hat{n}\partial_{\mu}\hat{n}\partial^{\mu}\hat{n}(\frac{1}{\partial^{2}+M^{2}})^{3}\Big]-4i\lambda_{\rm so}m^{3}{\rm Tr}\Big[\delta\hat{n}\partial_{\mu}n_{z}\partial^{\mu}\hat{n}(\frac{1}{\partial^{2}+M^{2}})^{3}\Big]-4i\lambda_{\rm so}m^{3}{\rm Tr}\Big[\delta\hat{n}n_{z}\partial_{\mu}\partial^{\mu}\hat{n}(\frac{1}{\partial^{2}+M^{2}})^{3}\Big]\notag\\
		&&-im^{4}\lambda_{\rm so}^{2}{\rm Tr}\Big[\delta\hat{n}\hat{n}n_{z}^{2}(\frac{1}{\partial^{2}+M^{2}})^{3}\Big]-im^{3}\lambda_{\rm so}^{3}{\rm Tr}\Big[\delta\hat{n}\sigma^z\hat{n}\sigma^{z}(\frac{1}{\partial^{2}+M^{2}})^{3}\Big]
		+i2m^{3}\lambda_{\rm so}{\rm Tr}\Big[\delta\hat{n}\sigma^z\partial_{\mu}\hat{n}\partial^{\mu}\hat{n}(\frac{1}{\partial^{2}+M^{2}})^{3}\Big]\\
		&=&-\frac{m^{4}}{16\pi M^{3}}\int dx^{3}\delta\hat{n}\hat{n}\partial_{\mu}\hat{n}\partial^{\mu}\hat{n}+\frac{\lambda_{\rm so}m^{3}}{8\pi M^{3}}\int dx^{3}\delta\hat{n}\partial_{\mu}n_{z}\partial^{\mu}\hat{n}+\frac{\lambda_{\rm so}m^{3}}{8\pi M^{3}}\int dx^{3}\delta\hat{n}n_{z}\partial_{\mu}\partial^{\mu}\hat{n}\notag\\
		&&+\frac{\lambda_{\rm so}^{2}m^{4}}{32\pi M^{3}}\int dx^{3}\delta\hat{n}\hat{n}n_{z}^{2}+\frac{\lambda_{\rm so}^{3}m^{3}}{32\pi M^{3}}\int dx^{3}\delta\hat{n}\sigma^z\hat{n}\sigma^{z}-\frac{\lambda_{\rm so}m^{3}}{16\pi M^{3}}\int dx^{3}\delta\hat{n}\sigma^z\partial_{\mu}\hat{n}\partial^{\mu}\hat{n}.\\
	\Eeq
	
	It can be clearly seen that the symmetry of the effective action is indeed reduced relative to the case of massless Dirac fermions due to the presence of $z$-directional kinetic terms containing $\sigma^{z}$. In general, the SOC produces a slightly different effect than simply adding a small mass $m_{0}$, they have the same effect on the topological term, but in the dynamical term, the SOC reduces the symmetry, while adding a small mass $m_{0}$ does not.

\section{SO(3) NLSM in forms of group variables}\label{app:eom}
The expression of the Hopf model is
\begin{align}
	S=\frac{1}{2\lambda}\int d^2 x{\rm Tr} (\partial_\mu \hat{n}\partial^\mu \hat{n})-\frac{2\pi}{24\pi^2}\int d^2 x{\rm Tr} (g^{-1}dg)^3,\label{F1}
\end{align}
where $g$ is the $SU(2)$ group element, $\hat{n}=\pmb{n}\cdot \pmb{\sigma}$, and $\pmb{n}$ is the unit vector in the 3D Euclidean space.Then the matrix form of $g$ is
\Beq
g=\left( \begin{array}{cc}
	z_1& -\bar {z_2}\\
	z_2& \bar {z_1}
\end{array} \right)
\Eeq
where $\det g=|z_1|^2+|z_2|^2=1$.To deal with two different field measure forms, we consider the concomitant representation of the $SU(2)$ group. $\hat{n}$ can be regarded as an element in the Lie algebra of the $SU(2)$ group. We can give the Lie algebra space the Killing-Cartan gauge ${\rm Tr} (\hat{n}\hat{n})=2\sum_{i=1}^{3}n_{i}^{2}$, then the Lie algebra space becomes a three-dimensional Euclidean space. We can represent $\hat{n}$ by the accompanying representation of the group,
\begin{align}\hat{n}=\pmb{n}\cdot \pmb{\sigma}= g\sigma^z g^{-1} \label{F2}
\end{align}

We can see that $g$ has three degrees of freedom (two complex numbers plus one restriction) while $\pmb n$ has only two (three real numbers plus one restriction). The map from $g$ to $\pmb n$ (or from $z$ to $\pmb n$)
is actually the Hopf mapping $S^{3}\rightarrow S^{2}$. Substitute Eq.(\ref{F2}) into the dynamical term in Eq.(\ref{F1}), one obtains
\begin{align}
	S_{\rm dyn}&=\frac{1}{2\lambda}\int d^2 x{\rm Tr}  (\partial_\mu \hat{n}\partial^\mu \hat{n})\notag\\
	&=\frac{1}{2\lambda}\int d^2 x{\rm Tr} \partial_\mu (g\sigma^zg^{-1})\partial^\mu (g\sigma^zg^{-1})\notag\\
	&=\frac{1}{\lambda}\int d^2 x{\rm Tr}  (\partial_\mu g\sigma^zg^{-1}+g\sigma^z\partial_\mu g^{-1})(\partial^\mu g\sigma^zg^{-1}+g\sigma^z\partial^\mu g^{-1})\notag\\
	&=\frac{1}{\lambda}\int d^2 x{\rm Tr}  (\partial_\mu g\partial^\mu g^{-1}+ g\sigma^z\partial_\mu g^{-1}g\sigma^z\partial^\mu g^{-1}).\label{F6}
\end{align}
\end{document}